\documentclass{article}
\usepackage[a4paper]{geometry}
\usepackage{wrapfig}
\usepackage{color}
\usepackage{xfrac}
\usepackage{pgfplots}
\usepackage{grffile}
\usepackage{graphicx}
\usepackage{float}
\usepackage[allbordercolors=black]{hyperref}
\usepackage{float}
\usepackage{eufrak}
\usepackage{mathrsfs}

\usepackage{tikz}
\usepgfplotslibrary{groupplots}
\usepgfplotslibrary{external} 
\pgfplotsset{compat=newest}
\usetikzlibrary{plotmarks}
\usetikzlibrary{patterns}
\usetikzlibrary{fit}

\usepackage{amsmath}
\usepackage{amssymb}

\usepackage{color}

\usepackage{caption}
\usepackage[final]{listofsymbols}
\usepackage{algorithm}
\usepackage{paralist}
\usepackage{float}

\definecolor{tudelft-cyan}{cmyk}{1,0,0,0}
\definecolor{tudelft-black}{cmyk}{0,0,0,1}
\definecolor{tudelft-white}{cmyk}{0,0,0,0}
\definecolor{tudelft-sea-green}{cmyk}{0.54,0,0.32,0}
\definecolor{tudelft-green}{cmyk}{1,0.15,0.4,0}
\definecolor{tudelft-dark-blue}{cmyk}{1,0.66,0,0.4}
\definecolor{tudelft-purple}{cmyk}{0.98,1,0,0.35}
\definecolor{tudelft-turquoise}{cmyk}{0.82,0,0.21,0.08}
\definecolor{tudelft-sky-blue}{cmyk}{0.45,0,0.06,0.06}
\definecolor{tudelft-lavendel}{cmyk}{0.45,0.2,0,0.07}
\definecolor{tudelft-orange}{cmyk}{0.02,0.56,0.84,0}
\definecolor{tudelft-warm-purple}{cmyk}{0.58,1,0,0.02}
\definecolor{tudelft-fuchsia}{cmyk}{0.19,1,0,0.19}
\definecolor{tudelft-bright-green}{cmyk}{0.36,0,1,0}
\definecolor{tudelft-yellow}{cmyk}{0.02,0,0.54,0}

\definecolor{armygreen}{rgb}{0.29, 0.33, 0.13}
\definecolor{beaver}{rgb}{0.62, 0.51, 0.44}
\definecolor{applegreen}{rgb}{0.55, 0.71, 0.0}
\definecolor{green-yellow}{rgb}{0.68, 1.0, 0.18}
\definecolor{inchworm}{rgb}{0.7, 0.93, 0.36}
\definecolor{lightblue}{rgb}{0.68, 0.85, 0.9}
\definecolor{lightkhaki}{rgb}{0.94, 0.9, 0.55}
\definecolor{ashgrey}{rgb}{0.7, 0.75, 0.71}
\definecolor{lightsalmon}{rgb}{1.0, 0.63, 0.48}

\def\wrmrk#1{{\color{tudelft-orange}[WIM: #1]}}
\def\wrmrk#1{}

\def\prmrk#1{{\color{tudelft-fuchsia}[PAWAN: #1]}}
\def\prmrk#1{}

\def\otext#1{{\color{red}#1}}
\def\revtext#1{{\color{black}#1}}

\def\btext#1{{\color{black}#1}}

\def\gtext#1{{\color{blue}#1}}

\usepackage{seismic_symbols_pawan}
\usepackage[numbers]{natbib}

\title
  {
A parameterization analysis for acoustic full-waveform 
inversion of sub-wavelength anomalies
}
\author
  {Pawan Bharadwaj$^1$, Wim Mulder$^{1,2}$ and Guy Drijkoningen$^1$ \\
  $^1$ Department of Geoscience \& Engineering, Delft University of Technology \\
  $^2$ Shell Global Solutions International B.V.  
  }
\begin{document}

\label{firstpage}

\maketitle

\begin{abstract}
With single-parameter full-waveform inversion (FWI), estimating the
inverse of the Hessian matrix will accelerate the convergence but 
	\revtext{will increase the computational cost and memory requirements 
	at each iteration.}
	%is computationally expensive.  
	Therefore, it is often replaced by the
inverse of an approximate Hessian that is easier to compute and serves
as a preconditioning matrix.
Alternatively, or in combination, a BFGS-type of optimization method
that estimates the inverse Hessian from subsequent iterations can be
applied.
Similarly, in the case of multi-parameter full-waveform inversion, the
	computation of the \revtext{additional} 
	Hessian terms that contain derivatives with respect
to more than one type of parameter %, called cross-parameter Hessian terms,
	is necessary.
%is not usually feasible at each iteration.  
%
If a simple gradient-based minimization with, for instance, just scalar
weights for each of the parameter types is used, different choices
of parameterization can be interpreted as different preconditioners
	\revtext{that change the condition number of the Hessian.}
If the non-linear inverse problem is well-posed, 
	then the inversion
should converge to a band-limited version of the true solution
irrespective of the parameterization choice, provided we start 
sufficiently close to the global minimum.
However, the choice of parameterization will affect the rate of
convergence to the exact solution and the `best' choice of
parameterization is the one with the fastest rate.

In this paper, we 
search for the best choice for acoustic full-waveform inversion,
where 
\begin{inparaenum}
\item anomalies with a size less than a quarter of the dominant wavelength 
	have to be estimated \revtext{without the risk 
	of converging to a local minimum};
\item the scattered wavefield is recorded at all the scattering angles;
\item \revtext{a steepest-descent minimization scheme is used.}
\end{inparaenum}
Towards that end, 
we review two different conventional analysis methods,
i.e., the point-scatterer analysis and diffraction-pattern analysis.
	\revtext{The conventional 
	methods consider only a few terms 
	of the full Hessian matrix for 
	parameterization analysis.}
To validate them, 
we consider single-component 
numerical examples,
where the inversion estimates one of the following:
\begin{inparaenum}
\item only \revtext{the} contrast of a point-shaped scatterer;
\item only \revtext{the} contrast of a Gaussian-shaped scatterer;
\item both \revtext{the} shape and contrast of a Gaussian-shaped scatterer.
\end{inparaenum}
\revtext{
The numerical examples show that 
the suggestions of the conventional analyses are only valid 
while estimating the contrast of point-shaped anomalies at a known location.
}
%
%On the other hand, 
For extended anomalies, the examples suggest that the best choice
of parameterization depends on the contrast of the subsurface
scatterer that the inversion tries to estimate.
%\revtext{which means the higher order terms of the 
%Hessian matrix cannot be neglected.}
%
Based on the results, we observe that there is no 
best parameterization choice for
full-waveform inversion in the case that
both the shape and the size of the anomalies have to be estimated.
We also
observe that a parameterization using the
acoustic impedance and mass density 
has the worst convergence rate.

In addition to full-waveform inversion, we consider 
Born modelling and inversion to learn 
if the dependence on the contrast of subsurface scatterer is due to the 
non-linearity in the full-waveform modelling and inversion.
We again observe that there is no best parameterization 
choice for Born modelling and inversion 
and the parameterization using the
acoustic impedance and mass density has the worst convergence
in the general case. 

\revtext{
Finally, 
we also show that the 
parameterization
analysis 
during a 
	hierarchical inversion, where 
	the data have limited scattering angles,
	only helps to 
	select a subspace for mono-parameter inversion.
	For multi-parameter hierarchical inversion,
the search for the \emph{best} parameterization 
in terms of
the 
convergence speed might be obfuscated by non-uniqueness problems.
}

\end{abstract}

%\begin{keywords}
% full-waveform inversion; parameterization; multi-parameter inversion.
%\end{keywords}

\newcommand{\logten}{\ensuremath{\log_{10}}}
\newcommand{\lxxrho}{$\mVEC_{\mB,\mR}$\xspace}
\newcommand{\velrho}{$\mVEC_{\mVp,\mR}$\xspace}
\newcommand{\ipxrho}{$\mVEC_{\mIp,\mR}$\xspace}
\newcommand{\velipx}{$\mVEC_{\mVp,\mIp}$\xspace}
\newcommand{\ilxirh}{$\mVEC_{\sfrac{1}{\mB},\sfrac{1}{\mR}}$\xspace}

\newcommand{\lxxrhop}{\lxxrho-parameterization\xspace}
\newcommand{\velrhop}{\velrho-parameterization\xspace}
\newcommand{\ipxrhop}{\ipxrho-parameterization\xspace}
\newcommand{\velipxp}{\velipx-parameterization\xspace}
\newcommand{\ilxirhp}{\ilxirh-parameterization\xspace}

\def\Wmod{\mathbf{W}}
\def\Hfull{\mathbf{H}}
\def\Hker{\mathbb{H}}

\section{Introduction}
Quantitative imaging of various near-surface elastic parameters is essential in many civil engineering applications as well as for hydrocarbon exploration.
One approach  is
full-waveform inversion (FWI) of the recorded elastic
wavefield \citep{ref:tarantola1986, ref:virieux2009}, which is
sensitive to the shear and compressional properties of the subsurface.
Multi-parameter FWI is a non-linear procedure that minimizes the least-squares misfit between the recorded and the modelled seismic data 
to estimate various subsurface parameters.
Given the size of the seismic problem, i.e., estimating thousands 
or millions of parameters, it is only feasible in practice to use descent methods for optimization.

In order to 
reduce the number of iterations needed to 
reach an acceptable solution, the 
gradient 
at each iteration 
\revtext{can} be preconditioned % should be preconditioned 
using the inverse of the Hessian matrix \citep{ref:pratt1998}.
The elements of the Hessian matrix are the second-order derivatives of the 
\revtext{data} misfit 
function with respect to the model parameters. 
If model parameter of kinds $a$ and $b$
at locations $\xx_i$ and $\xx_j$ are given by $m_a(\xx_i)$ and $m_b(\xx_j)$, respectively,
the elements of the Hessian matrix $\Hfull$ are % then given by
\begin{eqnarray}
  {\Hfull}_{(a,i),(b,j)} = {\partial_{m_a(\xx_i)}}
             {\partial_{m_b(\xx_j)} \funcls},
\label{eqn:param_Hfull}
\end{eqnarray}
where $\funcls$ denotes the least-squares misfit functional.
%

\begin{figure}
\begin{center}
	\includegraphics[width=0.5\textwidth]{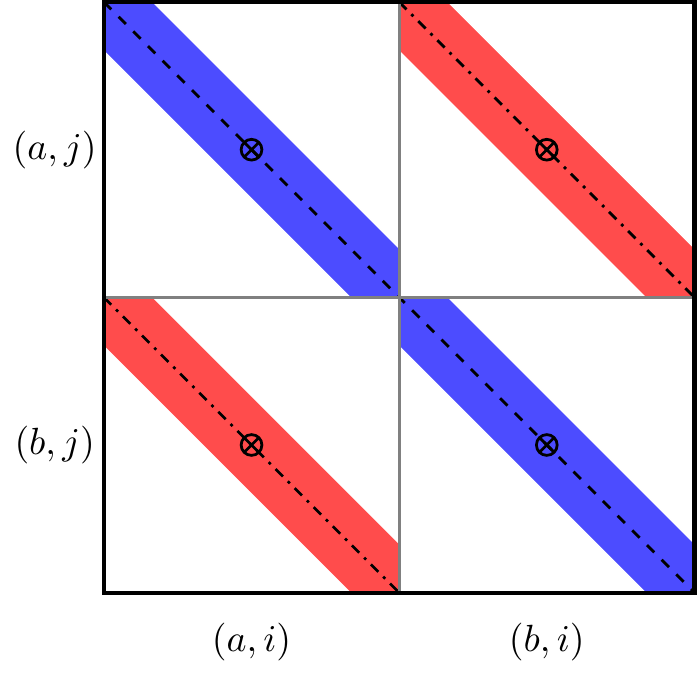}
%\begin{tikzpicture}
%
%\blockmatrix{6}{6}{he}{black}{white}{true}{black}{blue}{0.5}
%
%\end{tikzpicture}
\caption{This sketch marks different terms of the Hessian matrix 
in equation~\ref{eqn:param_Hfull}.
The main-diagonal mono-parameter terms are marked with a dashed line.
The band-diagonal mono-parameter terms are marked in blue and
the block cross-parameter terms in red.
Dashed-dotted line mark the main cross-parameter terms. 
The terms of the reduced Hessian, in equation~\ref{eqn:param_Hker0}, are marked with $\otimes$.
}
\label{fig:param_Hsketch}
\end{center}
\end{figure}

% mono-parameter
We will refer to the 
Hessian terms with $a = b$ as \emph{mono-parameter} terms.
The terms of the Hessian on its band diagonal with $i \approx j$, $i \ne j$  and $a = b$ 
will be called \emph{band-diagonal} mono-parameter terms, as sketched in Figure~\ref{fig:param_Hsketch}.
During the preconditioning, they
deconvolve 
the gradient such that 
it is less dependent on source bandwidth and acquisition geometry.
The terms of the Hessian 
where $i = j$ and $a = b$, called \emph{main-diagonal} mono-parameter terms, account for the amplitude effects
in wave propagation \citep{ref:pratt1998, ref:virieux2009},
for instance, due to geometrical spreading.
In the presence of 
multiple scattering, 
the change in data due to a model perturbation at $\xx_i$
depends on 
the model perturbation at $\xx_j$ even if $i \ne j$ and $i \not\approx j$.
This means that these 
non-band-diagonal 
terms of the Hessian,
${\Hfull}_{(a,i),(b,j)}$ with $i \neq j$ and $i \not\approx j$, 
are non-zero. 
During the preconditioning, 
they correct the gradient 
such that multiple scattering is taken into account.
%
%

% cross-parameter
The non-zero Hessian terms with $a \ne b$ 
represent cross-talk between 
the different parameter types.
We call them {\em{cross-parameter}} terms, 
\revtext{sketched} in Figure~\ref{fig:param_Hsketch}. 
The cross-parameter terms corresponding to a single subsurface position
when $i=j$ are called \emph{main} cross-parameter terms, whereas
the cross-parameter terms with 
$i \approx j$ and $i \ne j$ are called \emph{block} cross-parameter terms.
When cross-parameter terms are smallest, a change in the data due to a 
perturbation of one parameter should be almost independent of 
the perturbation of another parameter at the same location.

% master block matrix, should use wrappers instead of calling this
% directly
% This should be called within the tikzpicture environment
% \blockmatrix
%  {width}
%  {height}
%  {text}
%  {block_line_color} (can be none --> no line)
%  {block_fill_color} (can be none --> empty fill)
%  {is_diagonal} (true --> true, otherwise --> false)
%  {diagonal_line_color} (ignored if not diagonal) (can be none --> no line)
%  {diagonal_fill_color} (ignored if not diagonal) (can be noneo --> empty fill)
%  {diagonal_offset} (half diagonal width in tikz units)
% Note: colors here are not rgb, they are defined colors
\newcommand{\blockmatrix}[9]{
  \ifthenelse{\equal{#6}{true}}
  {
    \draw[draw=white,fill=blue!70] (0,#2) -- (#9,#2) -- (#1/2,#2/2+#9) -- ( #1/2,#2/2) -- ( #1/2 - #9,#2/2) -- (0,#2 -#9) -- cycle;
    \draw[draw=white,fill=blue!70] (#1/2,#2/2) -- (#1/2+ #9,#2/2) -- ( #1,#9) -- ( #1,0) -- ( #1 - #9,0) -- (#1/2,#2/2 -#9) -- cycle;
  }
  {}
  \ifthenelse{\equal{#6}{true}}
  {
    \draw[draw=white,fill=red!70] (0,#2/2) -- (#9,#2/2) -- ( #1/2,#9) -- ( #1/2,0) -- ( #1/2 - #9,0) -- (0,#2/2 -#9) -- cycle;
  }
  {}
  \ifthenelse{\equal{#6}{true}}
  {
    \draw[draw=white,fill=red!70] (#1/2,#2) -- (#1/2+#9,#2) -- ( #1,#2/2+#9) -- (#1, #2/2) -- ( #1-#9,#2/2 ) -- (#1/2,#2-#9) -- cycle;
  }
  {}
  \draw[thick,dashed, draw=#7] (0,#2) -- (#1,0);
  \draw[thick,dashdotted, draw=#7] (0,#2/2) -- (#1/2,0);
  \draw[thick,dashdotted, draw=#7] (#1/2,#2) -- (#1,#2/2);
  \draw[thick, draw=#7!50] (#1/2,0) -- (#1/2,#2);
  \draw[thick, draw=#7!50] (0,#2/2) -- (#1,#2/2);
  \draw[very thick,draw=#4] (0,0) rectangle( #1,#2);

\draw ( #1/4, 0-#9) node {$(a,i)$};
\draw ( #1*3/4, 0-#9) node {$(b,i)$};
\draw ( 0-#9,#2/4) node {$(b,j)$};
\draw ( 0-#9,#2*3/4) node {$(a,j)$};

\draw (#1/4,#2/4) node[mark size=3pt, thick]{\pgfuseplotmark{otimes}};
\draw (#1*3/4,#2/4) node[mark size=3pt, thick]{\pgfuseplotmark{otimes}};
\draw (#1*3/4,#2*3/4) node[mark size=3pt, thick]{\pgfuseplotmark{otimes}};
\draw (#1/4,#2*3/4) node[mark size=3pt, thick]{\pgfuseplotmark{otimes}};
}

% What is Hessian computation  is expensive

The computation of the Hessian matrix or its inverse 
will
increase the computational cost and memory requirements
at each iteration
%is computationally 
%expensive i
for large-scale \revtext{3-D} acoustic inverse problems 
\citep{ref:pratt1998, Fichtner2011, metivier2014full}. 
Therefore,
in this paper, we restrict ourselves to gradient-based optimisation 
without involving the Hessian.
Since the terms of the Hessian matrix depend on the choice of the 
subsurface parameterization in the case of
multi-parameter full-waveform inversion, 
the various  parameterization choices are not equivalent. 
If the inverse problem is well-posed,
the inversion should 
converge to 
a band-limited version of the true solution 
irrespective of the
parameterization choice, provided we start close to the global minimum.
The \emph{best} choice of parameterization is the one that
converges the fastest. % with the fastest convergence rate.
It should be noted that the rate
of convergence of a particular 
parameterization depends 
on the background subsurface model, without scatterers,
and the acquisition geometry.
The background subsurface model serves as the initial or starting model for inversion.

Hence, for a given acquisition geometry and 
background subsurface model, 
it is obvious that the fastest convergence
will be obtained for a parameterization choice 
that provides zero cross-parameter terms in the Hessian.
% with zero cross-parameter Hessian terms has the fastest convergence rate.
%
We notice,
for an acoustic inverse problem, 
that
both the main and block cross-parameter 
terms are non-zero for any 
of the parameterization choices 
involving wave velocity, density, bulk modulus or wave-impedance.
Among these options, we, therefore, have to choose a parameterization that provides the smallest condition number of the Hessian matrix.
\revtext{
Because the computation of the 
full Hessian matrix and its
condition number
for large-scale inverse problems is 
costly and requires large memory,
we prefer a simpler method for parameterization analysis.}
\prmrk{This is for the condition number of the Hessian, but not the matrix itself.}
\wrmrk{But there are fast approximate methods to estimate the condition number, such as the Jacobi-Davidson method! Matlab has {\texttt{condest}}.}

There are two conventional methods to analyse different subsurface 
parameterization choices with the aim to find the \emph{best} one. 
The first method, called diffraction-pattern analysis, chooses the best parameterization by examining the radiation patterns of the scattered wavefield generated by different perturbations \citep{ref:tarantola1986, operto2013guided, prieux2013multiparameter}. 
The second method, called point-scatterer analysis, considers the Hessian of a 
reduced inverse problem that only estimates the contrast of
a point scatterer at a known location. 
The reduction will result in fewer unknowns compared to the 
original inverse problem, where both the contrast and shape of the scatterer have to be estimated.
The block cross-parameter terms of the original non-reduced Hessian, the red terms in Figure~\ref{fig:param_Hsketch}, are absent in the reduced Hessian.
Therefore, this analysis ignores the block cross-parameter terms of the original non-reduced Hessian matrix and chooses a
parameterization which has main cross-parameter terms of the least possible magnitude.
\revtext{
For elastic waveform inversion, 
\cite{modrak2016choice} 
numerically 
demonstrated 
that the choice of material parameters is more complicated than the current literature suggests. % reflects.
}

% talk about almost well-posed problem used in this paper
% this problem is used to show inefficiency of 
% Approximate Hessian analysis and diffraction-pattern analysis.
In this paper, we consider an almost 
well-posed 2-D acoustic inversion example.
In the numerical examples, the aim is to 
reconstruct seven different scatterers surrounded by sources 
and receivers using the steepest-descent minimization method, 
which does not 
involve \revtext{any} computation of the Hessian matrix at each iteration.
The steepest-descent method, unlike the conjugate-gradient or L-BFGS quasi-Newton \citep{ref:lbfgs} methods \revtext{that are preferred in practice}, 
allows us to clearly observe the differences
in convergence rates among the various parameterization choices.
In this regard, we employ three different modelling and inversion schemes, where
the first two schemes use the Born approximation for modelling and inversion, 
while the third scheme
uses a time-domain acoustic staggered-grid finite-difference
code.
For Born modelling and inversion, we use a homogeneous background model 
and the analytical expressions of the corresponding 2-D Green functions.
With this example, 
we show that the findings of the conventional parameterization analysis methods 
are not valid in general.
%
%
% about error bowl analysis 
% 

This paper is organised as follows.
We start with % listing 
five different parameterization choices for acoustic inversion whose performance will be investigated.
Then, we outline the three modelling and inversion schemes that are employed in the numerical example.
Next, 
we briefly review the two conventional parameterization analysis methods before 
validating them against the numerical examples.
The last section concludes the paper.

\section{Parameterization Choices}

\begin{table}
\caption{Various medium parameterizations that are used in this paper for the acoustic inverse 
problem.}
\label{tab:param_param}
\begin{center}
\begin{tabular}{ | l | c | c | r |}
\hline
& Parameterization & First Parameter & Second Parameter \\ \hline
(A)&\velrho & $\xi_{\mVp}=(\mVp-\mVpo)/\mVpo$ & $\xi_{\mR}=(\mR-\mRo)/\mRo$  \\ \hline
(B)&\ipxrho & $\xi_{\mIp}=(\mIp-\mIpo)/\mIpo$ & $\xi_{\mR}=(\mR-\mRo)/\mRo$  \\ \hline
(C)&\velipx & $\xi_{\mVp}=(\mVp-\mVpo)/\mVpo$ & $\xi_{\mIp}=(\mIp-\mIpo)/\mIpo$  \\ \hline
(D)&\lxxrho & $\xi_{\mB}=(\mB-\mBo)/\mBo$ & $\xi_{\mR}=(\mR-\mRo)/\mRo$ \\\hline
(E)&\ilxirh & $\xi_{\sfrac{1}{\mB}}=(\sfrac{1}{\mB}-\sfrac{1}{\mBo})/(\sfrac{1}{\mBo})$ & $\xi_{\sfrac{1}{\mR}}=(\sfrac{1}{\mR}-\sfrac{1}{\mRo})/(\sfrac{1}{\mRo})$ \\\hline
\end{tabular}
\end{center}
\end{table}

We parameterize the Earth model with % using 
two parameters at each point, involving
density, $\mR$, and compressional-wave speed, $\mVp$, or their
combinations, for instance, compressional-wave impedance, $\mIp = \mR \mVp$, or
bulk modulus, $\mB = \mR \mVp^2$.
We denote the 
model parameter of kind $a$ by $m_a$ and 
write 
\begin{eqnarray}
m_a(\xx) = m_{ao}(\xx) \left[1+\xi_{m_a}(\xx)\right],
\end{eqnarray}
where $\xi_{m_a}(\xx)$
% = \left(m^{(i,j)}-m^{(i,j)}_o\right)\left(m^{(i,j)}_o\right)^{-1}$ 
is the contrast function and 
an additional subscript $o$ is used to denote the parameters corresponding to the reference medium.
The subsurface location is denoted by $\xx$.
When the subsurface is parameterized using $\xi_{m_a}$ and $\xi_{m_b}$,
the model vector at each point is denoted by  
$\mVEC_{m_a,m_b}$ and the choice of
parameterization is indicated by the subscript `$m_a,m_b$'.  
For example, 
in the case of \velrhop, the subsurface is 
parameterized using the contrast functions of the compressional-wave speed $\mVp$ and 
mass density $\mR$.
The 
model vector for \velrhop is  
\begin{eqnarray}
\mVEC_{\mV,\mR} = \begin{bmatrix}
         \xi_\mV \\ \xi_\mR
         \end{bmatrix}.
\end{eqnarray}
In this paper, we consider the five different 
parameterization choices given in 
the Table~\ref{tab:param_param}.
In the case of a \velipx-parameterization,
perturbations in the first 
subsurface parameter mainly  % \emph{mainly} 
affect the transmission of waves 
and perturbations in the second mainly their reflections.
We consider the \ilxirh-parameterization 
because the 
acoustic wave-operator is linear in $\xi_{\sfrac{1}{\mB}}$ and $\xi_{\sfrac{1}{\mR}}$, 
providing zero second derivatives with respect to these 
medium parameters. 
The second derivatives, which are non-zero for other parameterization 
choices, are used 
\revtext{in the formulae}
%during the computation 
of the full Hessian matrix as shown by \cite{Fichtner2011}.
\revtext{
	These are neglected during the conventional parameterization analysis methods.
	We consider the \lxxrhop, even though it has similar 
	first-order Hessian information as that of \ilxirhp, 
	to understand the importance of the second derivatives.
	In other words, 
	a significant difference in the convergence 
	rates of 
	\lxxrho- and \ilxirh-parameterizations will suggest that the 
	conventional analysis is not sufficient for multi-parameter FWI.
	}
Now, we discuss two approaches
to re-parameterize the subsurface from one parameterization to another.
\subsection{Re-parameterization}

\begin{table}
\caption{Conversion formulas used for the non-linear re-parameterization.}
\label{tab:param_con2}
\begin{center}
\begin{tabular}{ | l | c | c | r |}
\hline
& Parameterization & First Parameter & Second Parameter \\ \hline
(A)&\velrho & $\xi_{\mVp}=\sqrt{\frac{\left(\xi_{\sfrac{1}{\mR}} + 1\right)}{\left(\xi_{\sfrac{1}{\mB}} + 1\right)}}-1$ & $\xi_{\mR}=- \frac{\xi_{\sfrac{1}{\mR}}}{\xi_{\sfrac{1}{\mR}} + 1}$  \\ \hline
(B)&\ipxrho & $\xi_{\mIp}=\sqrt{\frac{1}{\left(\xi_{\sfrac{1}{\mR}} + 1\right)\left(\xi_{\sfrac{1}{\mB}} + 1\right)}}-1$ & $\xi_{\mR}=- \frac{\xi_{\sfrac{1}{\mR}}}{\xi_{\sfrac{1}{\mR}} + 1}$  \\ \hline
(C)&\velipx & $\xi_{\mVp}=\sqrt{\frac{\left(\xi_{\sfrac{1}{\mR}} + 1\right)}{\left(\xi_{\sfrac{1}{\mB}} + 1\right)}}-1$ & $\xi_{\mIp}=\sqrt{\frac{1}{\left(\xi_{\sfrac{1}{\mR}} + 1\right)\left(\xi_{\sfrac{1}{\mB}} + 1\right)}}-1$  \\ \hline
(D)&\lxxrho & $\xi_{\mB}=- \frac{\xi_{\sfrac{1}{\mB}}}{\xi_{\sfrac{1}{\mB}} + 1}$ & $\xi_{\mR}=- \frac{\xi_{\sfrac{1}{\mR}}}{\xi_{\sfrac{1}{\mR}} + 1}$ \\\hline
\end{tabular}
\end{center}
\end{table}

The non-linear re-parameterization approach involves the exact non-linear transformation.
For example, in order to obtain $\xi_{\mB}$ from $\xi_{\sfrac{1}{\mB}}$, we write 
\begin{eqnarray}
\xi_{\mB} = \frac{\mB - \mBo}{\mBo} = 
\left({\frac{1}{\xi_{\sfrac{1}{\mB}}\sfrac{1}{\mBo} + \sfrac{1}{\mBo}} - \mBo}\right)\frac{1}{\mBo}
={\frac{1}{\xi_{\sfrac{1}{\mB}} + 1} - 1} = - \frac{\xi_{\sfrac{1}{\mB}}}{\xi_{\sfrac{1}{\mB}} + 1}.
\end{eqnarray}
Similarly, Table~\ref{tab:param_con2} gives non-linear re-parameterization formulas for various parameterization choices.

\begin{table}
\caption{Conversion formulas used for the linear re-parameterization.}
\label{tab:param_con1}
\begin{center}
\begin{tabular}{ | l | c | c | r |}
\hline
& Parameterization & First Parameter & Second Parameter \\ \hline
(A)&\velrho & $\xi_{\mVp}=\tfrac{1}{2}\left(\xi_{\sfrac{1}{\mR}}-\xi_{\sfrac{1}{\mB}}\right)$ & $\xi_{\mR}=-\xi_{\sfrac{1}{\mR}}$  \\ \hline
(B)&\ipxrho & $\xi_{\mIp}=\xi_{\sfrac{1}{\mR}} - \tfrac{1}{2}\xi_{\sfrac{1}{\mB}}$ & $\xi_{\mR}=-\xi_{\sfrac{1}{\mR}}$  \\ \hline
(C)&\velipx & $\xi_{\mVp}=\tfrac{1}{2}\left(\xi_{\sfrac{1}{\mR}}-\xi_{\sfrac{1}{\mB}}\right)$ & $\xi_{\mIp}=\xi_{\sfrac{1}{\mR}} - \tfrac{1}{2}\xi_{\sfrac{1}{\mB}}$  \\ \hline
(D)&\lxxrho & $\xi_{\mB}=-\xi_{\sfrac{1}{\mB}}$ & $\xi_{\mR}=-\xi_{\sfrac{1}{\mR}}$ \\\hline
\end{tabular}
\end{center}
\end{table}

The linear re-parameterization approach assumes small 
contrasts,
ignoring higher-order terms in the exact non-linear conversion formula.
As an example, 
we consider a case where the subsurface is initially parameterized 
by \lxxrho.
In order to re-parameterize the subsurface to \ilxirh,
we write the model vector as
\begin{eqnarray}
\mVEC_{\sfrac{1}{\mB},\sfrac{1}{\mR}} = 
\begin{bmatrix}
\xi_{\sfrac{1}{\mB}} \\ \xi_{\sfrac{1}{\mR}}
\end{bmatrix} 
 & = &
\begin{bmatrix}
-\frac{\mB_o}{\mB} & 0 \\ 0 & -\frac{\mR_o}{\mR} 
\end{bmatrix}
\begin{bmatrix}
\xi_{\mB} \\ \xi_{\mR}
\end{bmatrix} \nonumber \\
& = &
\begin{bmatrix}
-1 & 0 \\ 0 & -1 
\end{bmatrix}
\begin{bmatrix}
\xi_{\mB} \\ \xi_{\mR}
\end{bmatrix}
+
\begin{bmatrix}
\frac{\xi_{\mB}}{1 + \xi_{\mB}} & 0 \\ 0 & \frac{\xi_{\mR}}{1+\xi_{\mR}}
\end{bmatrix}
\begin{bmatrix}
\xi_{\mB} \\ \xi_{\mR}
\end{bmatrix}.
\end{eqnarray}
Now, the second term
on the right-hand side of the above equation 
is ignored since it is of the order $\xi_{\mB}^2$ and $\xi_{\mR}^2$ for small $\xi_{\mB}$ and $\xi_{\mR}$.
The linear re-parameterization formula becomes
\begin{eqnarray}
\mVEC_{\sfrac{1}{\mB},\sfrac{1}{\mR}} \approx \begin{bmatrix}
 -1 & 0 \\ 0 & -1 
 \end{bmatrix}
\mVEC_{\mB,\mR} = -\mVEC_{\mB,\mR}.
\end{eqnarray}
Similarly, 
Table~\ref{tab:param_con1} gives linear re-parameterization formulas for other 
parameterization choices.

\section{Modelling and Inversion}

We denote the 
2-D spatial coordinates 
by $\xx = (\x, \z)$, the 
origin by $\xx_0 = (0,0)$ and the 
positions of sources and receivers 
by $\xxs$ and $\xxr$, respectively.
We introduce 
the reference Green function, $G_o(\xx, \afreq;~\xxs)$, 
satisfying the 2-D acoustic wave equation  
\begin{eqnarray}
\Lwav_oG_o = \delta(\xx-\xxs). 
\end{eqnarray}
Here, $\afreq$ denotes angular frequency.
The wave operator $\Lwav_o$ is given by 
\begin{eqnarray}
\Lwav_o = \tfrac{\afreq^2}{\mB_o} + \nabla \cdot \tfrac{1}{\mR_o}\nabla, 
\end{eqnarray}
where $\mB_o$ denotes the bulk modulus and $\mR_o$ the mass density 
of the reference medium.
The Green function corresponding to the actual inhomogeneous 
medium is denoted by $G(\xx, \afreq;~\xxs)$ and satisfies the wave equation 
\begin{eqnarray}
\Lwav \, G = \delta(\xx-\xxs), 
\label{eqn:param_full}
\end{eqnarray}
with
\begin{eqnarray}
\Lwav = \tfrac{\afreq^2}{\mB} + \nabla \cdot \tfrac{1}{\mR}\nabla. 
\end{eqnarray}
We write Green's function in the actual inhomogeneous medium as $G = G_o + G_s$, where 
$G_s$ stands for the scattered component of the total pressure wavefield.
The Lippmann-Schwinger equation produces
\begin{eqnarray}
G(\xxr, \afreq;~\xxs) &=& G_o(\xxr, \afreq;~\xxs) + \nonumber \\
& & \int_{\xx} G_o(\xxr, \afreq;~\xx) \afreq^2\left(\tfrac{1}{\mB}-\tfrac{1}{\mBo}\right) G(\xx, \afreq;~\xxs) \diff\xx +  \nonumber \\
& & \int_{\xx} G_o(\xxr, \afreq;~\xx) \left(\nabla \cdot \left(\tfrac{1}{\mR}-\tfrac{1}{\mRo}\right)\nabla G(\xx, \afreq;~\xxs)\right) \diff\xx,
\label{eqn:param_lipp}
\end{eqnarray}
where $G_o(\xxr, \afreq;~\xx)$ denotes the scatterer-to-receiver field in the unperturbed medium and 
$G(\xx, \afreq;~\xxs)$ denotes the source-to-scatterer field in the perturbed medium.
We now employ the Born approximation to 
replace the actual Green function $G$ in the right-hand side of equation~\ref{eqn:param_lipp} 
with $G_o$.
Also, we use integration by parts to obtain
the Born approximation, $G_b$, of Green's function as
\begin{eqnarray}
G_b(\xxr, \afreq;~\xxs)  & = & G_o(\xxr, \afreq;~\xxs) +\nonumber \\
& &  \int_{\xx} \tfrac{\afreq^2}{\mBo} G_o(\xxr, \afreq;~\xxs) G_o(\xxr, \afreq;~\xxs)\xi_{\sfrac{1}{\mB}} \diff\xx - \nonumber \\ 
               &   &  \int_{\xx} \tfrac{1}{\mRo}\left[\nabla G_o(\xxr, \afreq;~\xxs)
         \cdot \nabla G_o(\xxr, \afreq;~\xxs)\right] \xi_{\sfrac{1}{\mR}} \diff\xx, \\
&=& G_o(\xxr, \afreq;~\xxs) + \int_{\xx} \Wmod^{\mathrm{T}}_{\sfrac{1}{\mB},\sfrac{1}{\mR}}(\xx, \afreq;~\xxr, \xxs)\,\mVEC_{\sfrac{1}{\mB},\sfrac{1}{\mR}}(\xx) \diff\xx,
\label{eqn:param_born}
\end{eqnarray}
where the modelling vector $\Wmod_{\sfrac{1}{\mB},\sfrac{1}{\mR}}$ is given by
\begin{eqnarray}
\Wmod_{\sfrac{1}{\mB},\sfrac{1}{\mR}}(\xx, \afreq;~\xxr, \xxs)  =  
\begin{bmatrix}
\tfrac{\afreq^2}{\mBo} G_o(\xxr,\xx) G_o(\xx, \xxs) \qquad -\tfrac{1}{\mRo}\left(\nabla G_o(\xxr, \xx)\right) \cdot \left(\nabla G_o(\xx,\xxs)   \right)
\end{bmatrix}^{\mathrm{T}}.
\end{eqnarray}

We now discuss different modelling and 
inversion schemes that are used in the numerical experiments in this paper.
In each scheme, 
the least-squares misfit between 
the artificially-generated \emph{observed} data $\qfreq(\xxr, \afreq;~\xxs)$
and the modelled data $\pfreq(\xxr, \afreq;~\xxs)$, 
\begin{eqnarray}
\funcls = \frac{1}{2}\sum_{\afreq \ge 0} \sum_{s, r}\|\pfreq(\xxr, \afreq;~\xxs)-\qfreq(\xxr, \afreq;~\xxs)\|^2,
\end{eqnarray}
is minimized.
If the subsurface is parameterized by a model vector other than
\ilxirh, either the linear or the non-linear re-parameterization approach is
used.
Then, 
for forward and adjoint modelling, 
we use either the Born or a full-waveform approach.
After the adjoint modelling, 
the gradient of $\funcls$ with respect to
\ilxirh is obtained by chain rule.
As already mentioned, we use a steepest-descent algorithm to minimize $\funcls$.
\begin{figure}
\begin{center}
	\includegraphics[width=0.8\textwidth]{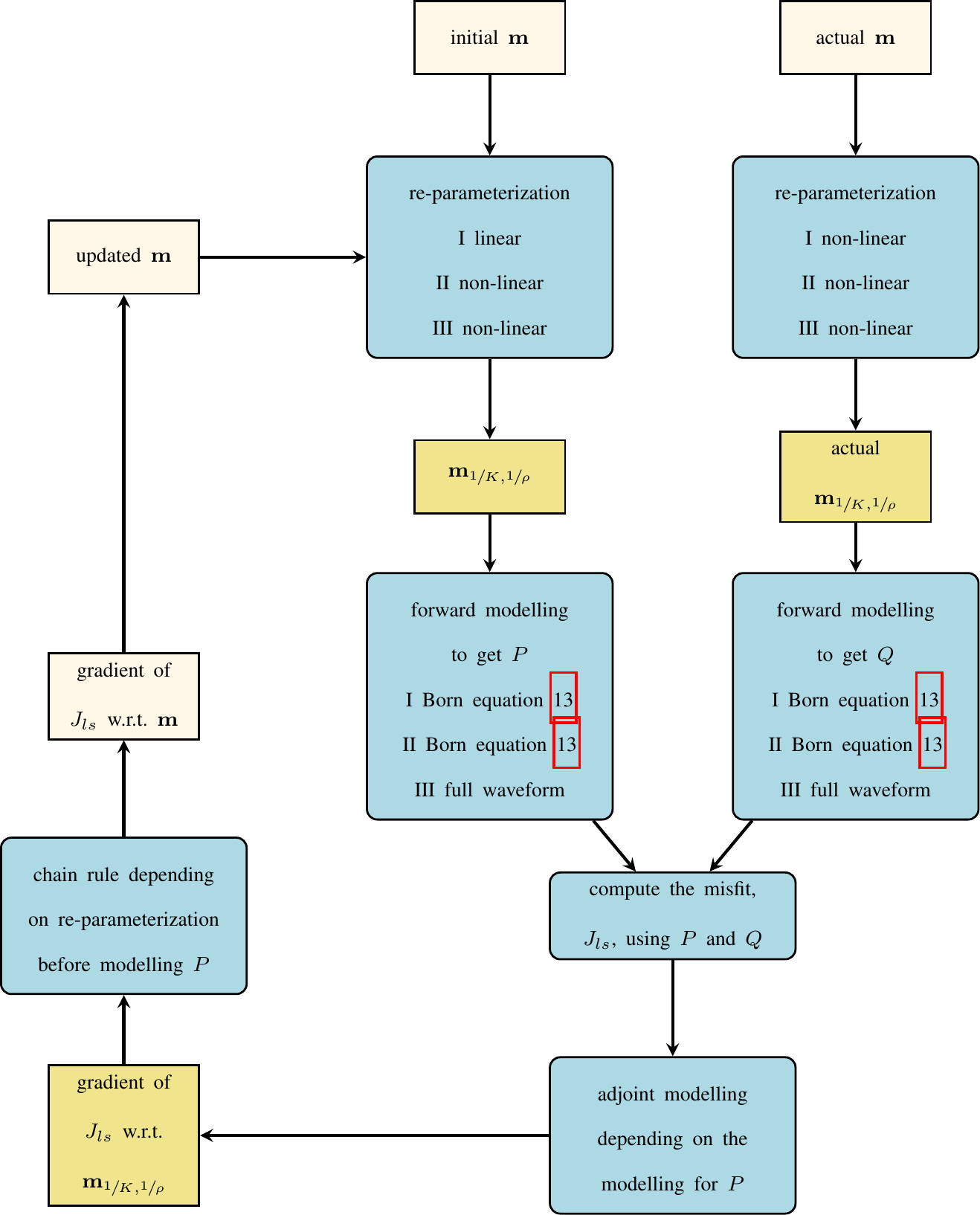}
\caption{
A flowchart to illustrate the modelling and inversion schemes I, II and III. 
Square boxes represent subsurface models and boxes with rounded corners represent an operation.
Boxes that depend on the parameterization choice are white.
Linear and non-linear re-parameterizations are described by the formulas in Tables~\ref{tab:param_con1} and \ref{tab:param_con2}, respectively.
}
\label{fig:param_scheme1}
\end{center}
\end{figure}

\subsection{Scheme I}
With this scheme, 
the modelled data are obtained by using 
the linear re-parameterization, listed in Table~\ref{tab:param_con1},
and the Born-approximated Green functions $G_b$ 
of equation~\ref{eqn:param_born}:
\begin{eqnarray}
\pfreq(\xxr, \afreq;~\xxs) = \sfofreq(\afreq)G_b(\xxr, \afreq;~\xxs).
\label{eqn:param_pfreq1}
\end{eqnarray}
Here, $\sfofreq(\afreq)$ denotes 
the source spectrum, which is assumed to be known.
Also, the `observed' data are generated with $G_b$, 
committing an inverse crime, but for those,
the non-linear re-parameterization approach of Table~\ref{tab:param_con2}
is used. 
The flowchart in Figure~\ref{fig:param_scheme1} 
illustrates this scheme.

\subsection{Scheme II}
As shown in Figure~\ref{fig:param_scheme1}, this scheme is similar to the modelling and inversion scheme I,
except that the non-linear re-parameterization, listed in Table~\ref{tab:param_con2}, is adopted 
while generating both the modelled and the `observed' data.

\subsection{Scheme III}

The numerical results obtained for this  
modelling and inversion scheme are decisive because it
is identical to 
conventional acoustic full-waveform modelling and inversion.
We used a time-domain acoustic staggered-grid finite-difference
code to solve equation~\ref{eqn:param_full} 
for the forward as well as the 
adjoint wavefield computations required for the gradients \citep{ref:tarantolaacou,ref:tarantola1986}.
Absorbing boundary conditions are used on all sides of the computational domain.
The flowchart in the Figure~\ref{fig:param_scheme1} also illustrates this scheme.

%#######################################################################
\section{Diffraction-pattern Analysis}

\begin{figure}
\begin{center}
	\includegraphics[width=0.8\textwidth]{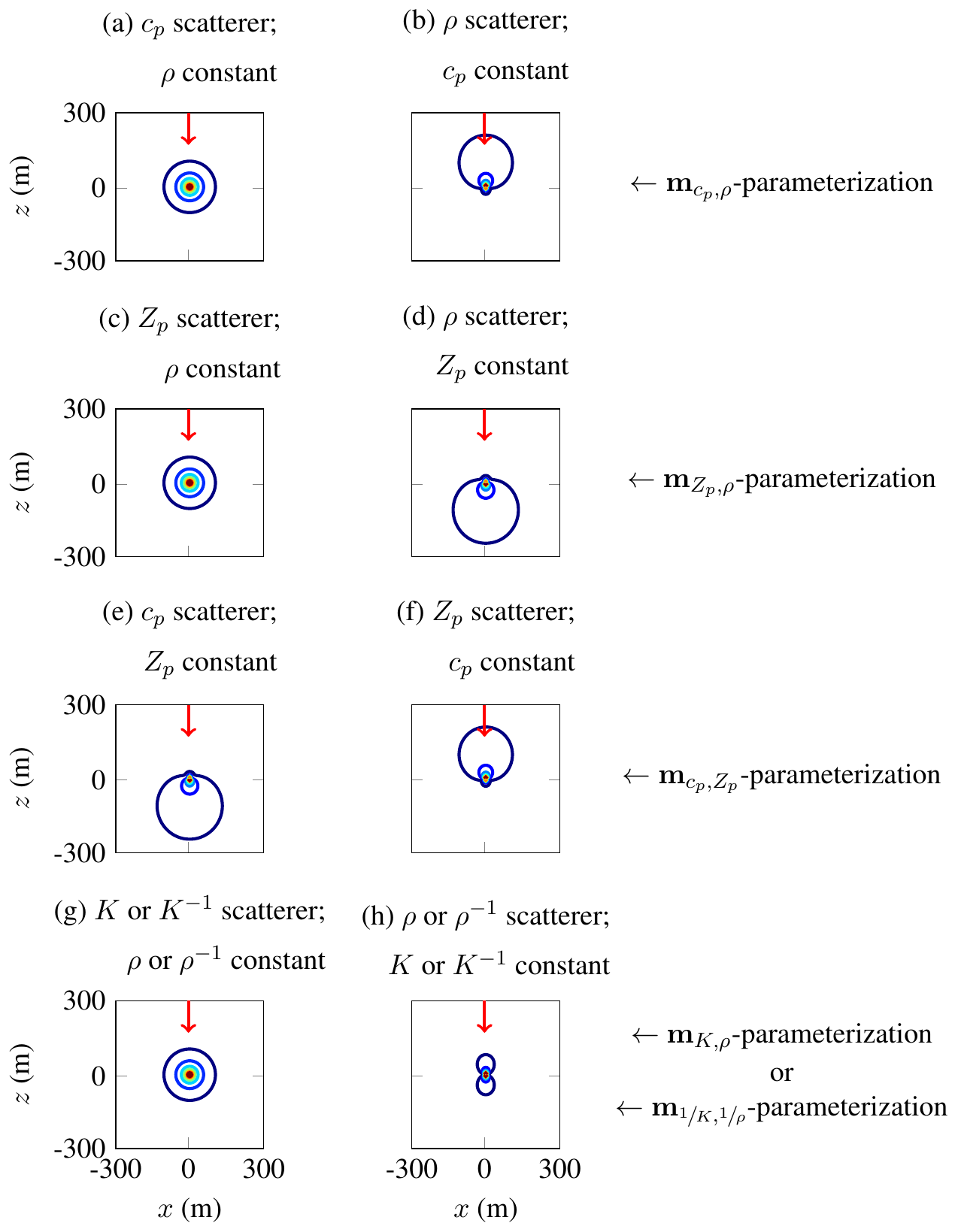}
\end{center}
\caption{
Diffraction patterns due to different point scatterers located at the center of the model
(0,0).
The red arrow indicates the direction of the incident primary wavefield radiated by a source located outside the plotted domain at (0,490).
Point scatterers have (a) velocity and (b) density perturbations for 
a \velrhop. 
(c), (d) For a \ipxrhop. (e), (f) For a \velipxp. (g), (h) For a \lxxrhop 
 or \ilxirhp. Red-coloured contours are used for higher scattered energy than blue-coloured contours.
}
\label{fig:param_radiation}
\end{figure}

The most common way to analyse various parameterization 
choices is with diffraction patterns \citep{ref:WuAki,ref:tarantola1986,ref:Malinowski,operto2013guided,ref:Gholami1,prieux2013multiparameter,He2016}.
In this analysis, 
for each individual parameterization choice, 
a perturbation of both the first and the second parameter
in the same scatter point at the centre, $\xx_0 = (0,0)$,
of the model is considered, leading to an inverse problem with two
parameters.
When the primary wavefield, incident on these perturbations,
is scattered, the contours of the scattered energy produce
diffraction patterns.
For example, in the case of a \velrhop, contours of the scattered energy 
due to a point scatterer with velocity and density contrasts are plotted in Figures~\ref{fig:param_radiation}a and \ref{fig:param_radiation}b, 
respectively.
Here, the incident wavefield is generated by a point source at $(0,490)$.
Note that these plots are insensitive to the sign of the incident wavefield.
The diffraction patterns for other parameterization choices are also plotted in Figure~\ref{fig:param_radiation}.
Alternatively, analytical expressions
derived in the framework of asymptotic ray+Born inversion \citep{forgues1997parameterization} can be used to obtain the diffraction patterns.

Diffraction pattern analysis
chooses subsurface parameterization
depending on the recorded arrivals, at a particular scattering angle $\theta$, which are being inverted or fit.
A parameterization where the two 
diffraction patterns at $\theta$
differ as much as possible 
has the fastest convergence because 
the change
in the data due to a 
perturbation in the first parameter is independent
of that caused by the second parameter.
For example, when mainly inverting arrivals recorded at 
short-to-intermediate scattering angles, 
the
trade-off between the two parameters is minimum in the case of 
\begin{inparaenum}
\item \velipxp\,--- $\mR$ doesn't affect the data; 
\item \ipxrhop\,--- $\mVp$ doesn't affect the data,
\end{inparaenum}
as shown 
in the Figures~\ref{fig:param_radiation}c, \ref{fig:param_radiation}d, 
\ref{fig:param_radiation}e and \ref{fig:param_radiation}f.
This is due to the fact that there is an 
overlap in the diffraction patterns in the case of other parameterization choices
(Figures~\ref{fig:param_radiation}a, \ref{fig:param_radiation}b, \ref{fig:param_radiation}g
and \ref{fig:param_radiation}h).
\revtext{
	Therefore,
	as shown by many authors,
	this analysis is useful
	to select the subspace for 
	mono-parameter inversion
	in
	separate scattering regimes.
	}
%	therefore the solution to the multi-parameter inverse problem cannot be determined uniquely.
%	%
%	However, 
%	%
%	We discuss this in the section~\ref{sec:nonuni}, where we show that 
%	the
%parameterization analysis is meaningless for inversion in regimes with 
%limited scattering angles.
%
\revtext{
However, in this paper, we focus
on the case 
in which the recorded 
arrivals for all the scattering angles are considered simultaneously. 
In this case,
the contrasts in both parameters can be 
uniquely determined as the inverse problem 
is almost well-posed.
}
The diffraction-pattern 
analysis suggests that a
\velipxp  is the \emph{best} choice for an almost well-posed problem because
its two diffraction patterns 
differ the most. 
\wrmrk{But my asymptotic analysis suggests 1/K and 1/rho?}
\prmrk{That is in the next section where the exact integration over all the angles is 
considered. When analyzing radiation patterns, perhaps we consider most 
of the scattering angles, but not all? 
This was also the recommended parameterization from 
Gholami et al. 2013.
To me, radiation-pattern analysis looks more qualitative, where we look at the 
shapes of squared energy contours (Figure 3)
without doing exact summations.
}

Later in this paper, we want to numerically validate this claim.
In the next section, we plot the terms of the Hessian matrix to
show that there is a dependence between the 
two parameters for all the parameterization choices, given that all the scattering angles are considered for inversion.

%#######################################################################
\section{Point-scatterer Analysis}

During this analysis,  the inverse problem is simplified 
such that only the 
contrast of a point-shaped scatterer at a known location $\xx_0$ has to be estimated, in our case $\xx_0=(0,0)$. 
In this case, the terms of the original Hessian matrix in equation~\ref{eqn:param_Hfull} are 
reduced to a 2 by 2 symmetric matrix since there are only two unknowns 
during the inversion:
\begin{eqnarray}
\Hker_{a,b} &=&  \begin{bmatrix} {\Hfull}_{(a,0),(a,0)} & {\Hfull}_{(a,0),(b,0)} \\ {\Hfull}_{(b,0),(a,0)} & {\Hfull}_{(b,0),(b,0)}  \end{bmatrix},
\label{eqn:param_Hker0}
\end{eqnarray}
where $\Hfull$ was defined in equation~\ref{eqn:param_Hfull}. 
We call the matrix $\Hker$ in equation~\ref{eqn:param_Hker0} 
the reduced Hessian. 
It varies with the parameterization choice and acquisition geometry.
As an example, 
when the modelling and inversion scheme I is employed and the 
subsurface is parameterized using \ilxirh, we 
can write $\Hker$ in terms of the modelling vector $\Wmod$ as
\wrmrk{ $W^\ast W^T$? I always forget which one it should be.}
\prmrk{You are right Wim, I defined $b=A^{T}x$ in eq.13, so, Hessian should be $A^{\ast}A^T$}
\begin{eqnarray}
\Hker_{\sfrac{1}{\mB},\sfrac{1}{\mR}} &=&   
	\sum_{\afreq} \sfofreq(\afreq)\sfofreq^{*}(\afreq)~\sum_{s, r} \Wmod^{\ast}_{\sfrac{1}{\mB},\sfrac{1}{\mR}}(\xx_0, \afreq;~\xxr, \xxs)\,\Wmod_{\sfrac{1}{\mB},\sfrac{1}{\mR}}^{\mathrm{T}}(\xx_0, \afreq;~\xxr, \xxs).
\label{eqn:param_Hker}
\end{eqnarray}

The point-scatterer analysis compares the reduced Hessians $\Hker$ for different parameterization choices with the objective
to find the one with the fastest convergence.
As the reduced Hessian also depends on the acquisition geometry, 
we choose, as an example, 
a circular acquisition geometry with sources and receivers along a circle with 
$\xx_0$ as the centre.
The eigenvalues and eigenvectors of the 
reduced Hessians \citep{plessix2011parametrization, operto2013guided} for the various parameterization choices 
can be plotted to assess the relative convergence rate for each choice.
% comment on the relative convergence rate of each choice.
%
This analysis was also 
employed for 
ray-based inversion by \cite{forgues1997parameterization},
where under the high-frequency approximation and 
in the absence of multiple scattering,
the terms of the non-reduced Hessian matrix, ${\Hfull}_{(a,i),(b,j)}$, are non-zero
only if $i=j$.
This analysis suggests that 
the \ilxirh and \lxxrho parameterization choices 
are equivalent and have the fastest convergence for the chosen circular acquisition geometry
for the following reasons:
\begin{itemize}
\item[\it{Condition Number}.]
  Convergence to the exact solution is possible in only one
 iteration step if all the eigenvalues of the reduced Hessian are equal. 
 Intuitively, this corresponds to the case when the ellipsoidal 
  contours of the objective function become circular.
 The condition number of $\Hker_{\sfrac{1}{\mB},\sfrac{1}{\mR}}$ or  $\Hker_{\mB,\mR}$, 
 unlike the reduced Hessians for other parameterization choices, is one.
\item[\it{Cross-parameter Terms}.] The convergence is faster
 when the coordinate axes, which refer to the parameters,  
        coincide with the eigenvector directions of the reduced Hessian. 
 This corresponds to the case where
        the cross-parameter terms of the reduced Hessian are zero. This is true only 
 if the subsurface is parameterized using either \lxxrho or \ilxirh.
\end{itemize}

\begin{figure}
\begin{center}
	\includegraphics[width=0.6\textwidth]{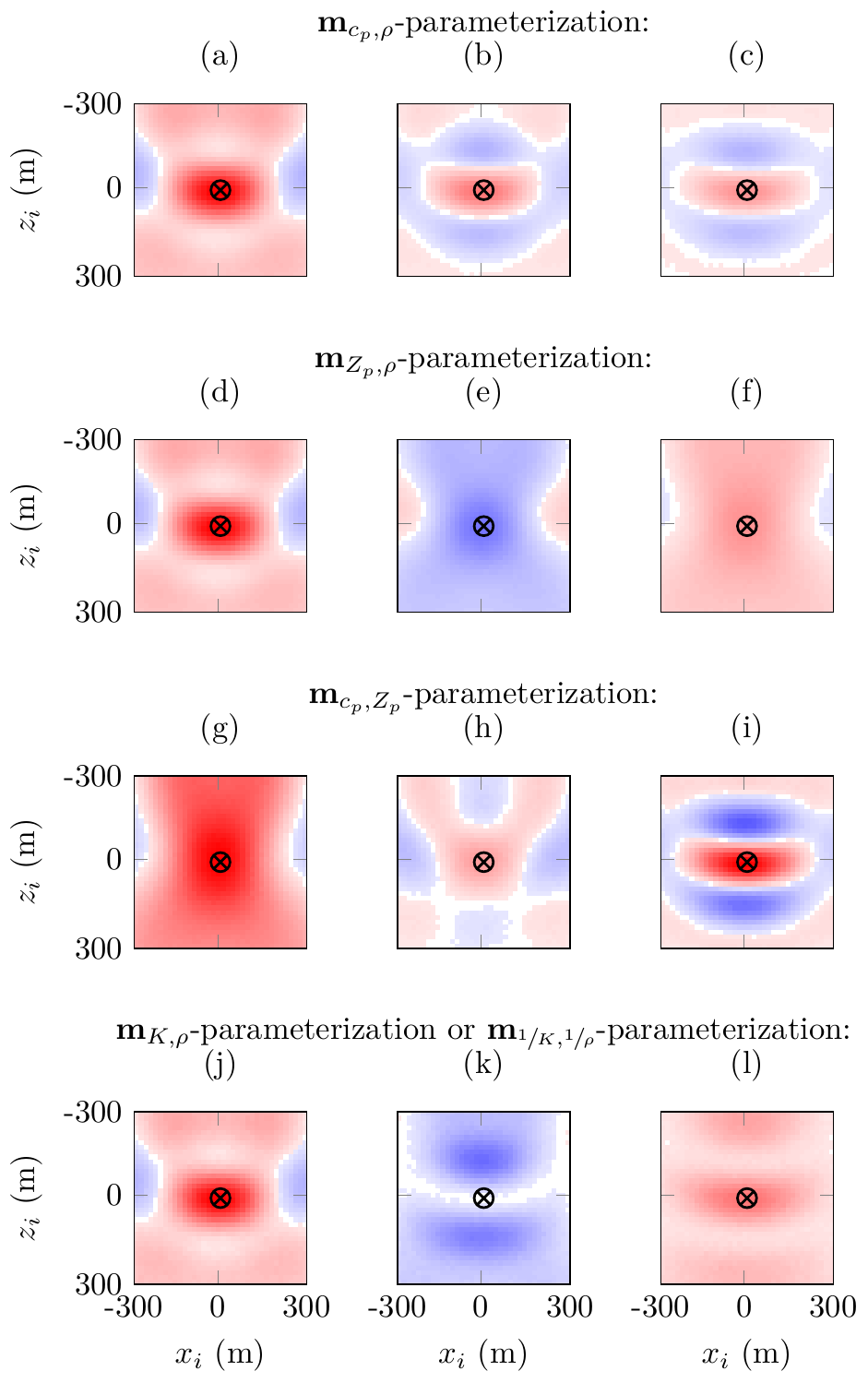}
\end{center}
\caption{
The terms of the non-reduced Hessian matrix, ${\Hfull}_{(a,i),(b,j)}$ such that $\xx_j = \xx_0$, plotted as 
a function of $\xx_i=(\x_i,\z_i)$ for the chosen, almost well-posed problem.
The modelling and inversion scheme I is adopted. A source is located at (0,$-$490) and receivers surround $\xx_0=(0,0)$ 
along a circle of radius 490$\,$m.
Each row shows the sub-matrices of $\Hfull$; the first and third columns 
show the mono-parameter terms while the second 
column shows the cross-parameter terms, respectively.
In ${\Hfull}_{(a,i),(b,j)}$, $a$ and $b$ are chosen as:
a) $\mVp$ and  $\mVp$;
b) $\mVp$ and  $\mR $;
c) $\mR $ and  $\mR $;
d) $\mIp$ and  $\mIp$;
e) $\mIp$ and  $\mR $;
f) $\mR $ and  $\mR $;
g) $\mVp$ and  $\mVp$;
h) $\mVp$ and  $\mIp$;
i) $\mIp$ and  $\mIp$;
j) $\mB$ and  $\mB$ or $\sfrac{1}{\mB}$ and $\sfrac{1}{\mB}$;
k) $\mB$ and  $\mR$ or $\sfrac{1}{\mB}$ and $\sfrac{1}{\mR}$;
l) $\mR$ and  $\mR$ or $\sfrac{1}{\mR}$ and $\sfrac{1}{\mR}$.
Observe that the 
block cross-parameter terms,
	plotted in the second column except at $(0,0)$, 
	of the non-reduced Hessian
are non-zero for any given parameterization choice.
	%, showing a dependence between the two parameters.
%
For each parameterization choice, the terms of the reduced Hessian used for the point-scatterer analysis are marked with $\otimes$.
%It can be seen that in the case of the \lxxrho- or \ilxirh-parameterization (j--k), the cross-parameter terms of the reduced Hessian are zero.
In this figure, red and blue colours represent positive and negative values, respectively.
White corresponds to values close to zero.
%
%For this reason, the point-scatterer analysis suggestions 
%that the \lxxrho- or \ilxirh-parameterization has the best convergence rate.
%
%
}
\label{fig:param_hess}
\end{figure}

% this analysis used in FWI
This analysis is not suited for realistic inverse problems, where both the shape and contrast of sub-wavelength scatterers have to be estimated, unlike the simplified one we considered with the point-shaped scatterer.
To show this, we employ the modelling and inversion scheme I and 
plot some terms of the non-reduced Hessian matrix, ${\Hfull}_{(a,i),(b,j)}$ with $\xx_j = \xx_0$, in Figure~\ref{fig:param_hess}.
Note that $\Hfull$ is a
normal matrix and therefore has three unknown 
sub-matrices since 
${\Hfull}_{(a,i),(b,j)} = {\Hfull}_{(b,i),(a,j)}$ for any given $i$ and $j$.
For this reason, only three plots are shown for each parameterization in Figure~\ref{fig:param_hess}.
It can be observed that the cross-parameter terms of the non-reduced Hessian matrix are non-zero for any given parameterization 
demonstrating the dependence between the 
two parameters.
As discussed before, the cross-parameter terms of the reduced Hessian,
i.e., the main cross-parameter terms of the non-reduced Hessian matrix,  
are zero (white colour in Figure~\ref{fig:param_hess}k) for a \lxxrho- or \ilxirh-parameterization.
However, 
the block cross-parameter terms are non-zero 
for these parameterization choices, as in 
Figure~\ref{fig:param_hess}k, resulting in the dependence between an $\mB$ or $\sfrac{1}{\mB}$ contrast at $\xx_0$ and 
a $\mR$ or $\sfrac{1}{\mR}$ contrast at any of the points neighbouring $\xx_0$.
Hence, there is a dependency between the parameters for all the parameterization choices.

Anyway, our goal is to validate the suggestions
of this analysis with the numerical examples later in this paper.

%#######################################################################
\section{Almost Well-posed Example}

\begin{figure}
\begin{center}
	\includegraphics[width=0.5\textwidth]{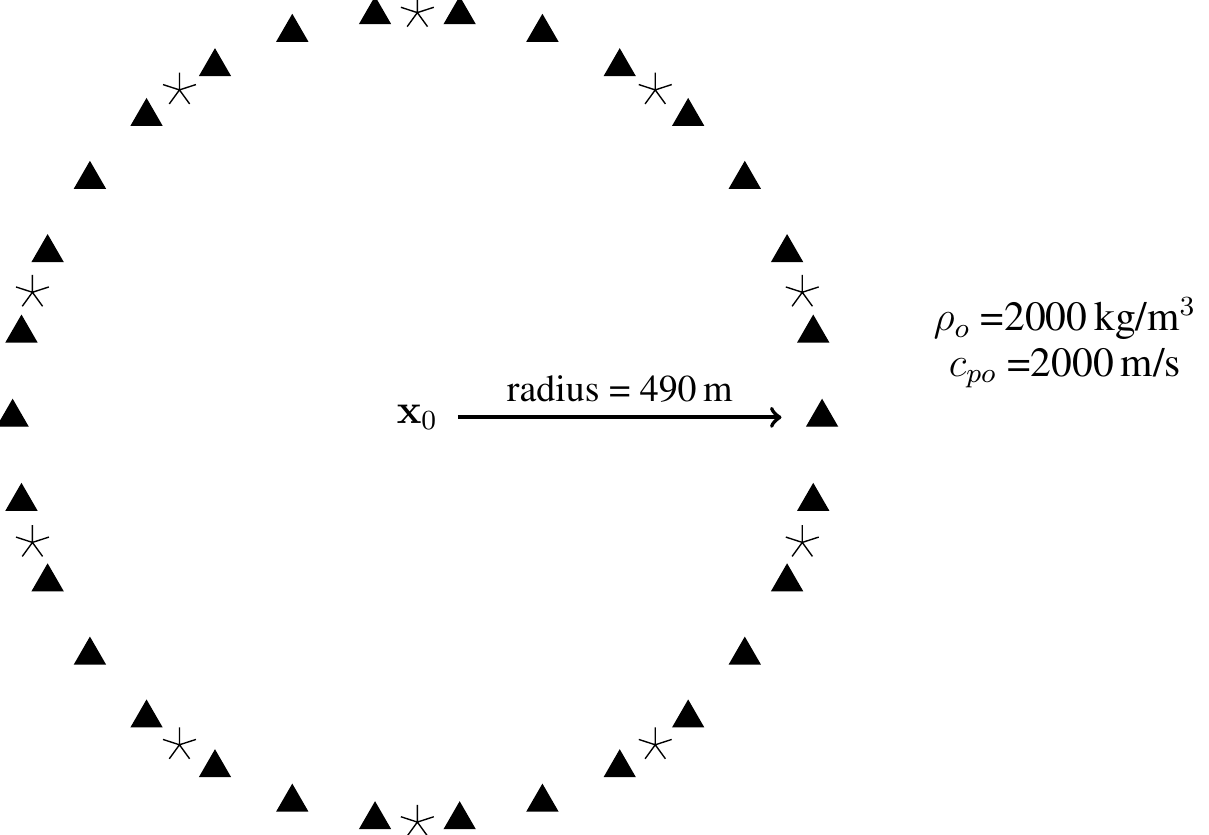}
%\begin{tikzpicture}
%\foreach \a in {1,2,...,30}{
%\draw (\a*360/30: 0.25\textwidth) node[mark size=5pt]{\pgfuseplotmark{triangle*}};
%}
%\foreach \b in {1,2,...,10}{
%\draw (\b*360/10+90: 0.25\textwidth) node[mark size=5pt]{\pgfuseplotmark{star}};
%}
%\draw(0,0) node{{\large$\xx_0$}};
%\draw(.4*\textwidth, 0.5) node{{\large \mVpo=2000$\,$m/s}};
%\draw(.4*\textwidth, 1.0) node{{\large \mRo=2000$\,$kg/m$^{3}$}};
%\draw[->,very thick] (0.025\textwidth,0) -- (0.25\textwidth-0.025\textwidth,0) node [above,midway] {radius = 490$\,$m};
%\end{tikzpicture}
\end{center}
\caption{
A sketch 
illustrating 
the almost well-posed example used for parameterization analysis.
% Stars mark the sources, triangles the receivers.
% We mark a source with a star and a receiver with a triangle.
We place 10 sources, marked by stars, and 100 receivers, marked by triangles, on a circle with a 490-m radius. 
Different 
scatterers listed in the Table~\ref{tab:param_scat} are
located at $\xx_0$.
}
\label{fig:param_problem}
\end{figure}

% explain well-posed problem
We define an almost well-posed numerical example 
by placing 10 sources and 100 receivers
evenly on a circle with a 490-m radius 
all around $\xx_0$
to avoid ill-posedness problems 
related to incomplete
illumination and angle-dependent information.
Figure~\ref{fig:param_problem} shows the setup.
\revtext{
	In a realistic setting, with one-sided acquisition, 
	a complete illumination of the target can never be achieved.
	This means, it is difficult 
	to 
	estimate the
	shape of the target
	without additional constraints.
	However, since 
	a wide-angle one-sided acquisition 
	can record
	all the angle-dependent information of the target,
	the inverse problem of estimating only the 
	contrast is almost well posed.
	\wrmrk{I do not understand how one-sided can replace a circle. They are essentially different.}
	\prmrk{I wanted to the say that the one-sided acq. 
	is same as circle acq.
	when the geometrical spreading is ignored and
	the aperture tends to infinity.
	Are there any other factors that should be taken care of?
	}
        \wrmrk{As an extreme example, take a nearly homogeneous medium. In transmission, you can estimate the velocity and attenuation, in reflection, there is nothing, there may be only a direct wave. So one-sided and circular are different.}
	\prmrk{When there is a source at $(-\infty, 0)$ and receiver at 
	$(+\infty,0)$, we are recording a transmitted waves even in the case of one-sided
	situation. So what I wanted to say is that with one-sided acquisition 
	we can also record all scattering angles from 0 to 180 degrees, just like circular acquisition. 
	But in a circular acquisition, we are illuminating the target point from all
	the source angles. For example, in one-sided acquisition 
	we cannot illuminate the target along $x=0$ and then record scattered energy at 180 degrees.
	If we are only interested in quantitative imaging, i.e., the shape of scattering in known, then we have all the scattering information in one sided acq. with wide aperture as that of a circular aperture.
	I have adjusted the text now.}
	%
	%Assuming a constant background velocity,
	%this acquisition will introduce
	%differences in the geometrical spreading 
	%among the scattering angles.
	%%
	%Hence, the translation implies that 
	%the scattered data at
	%all the angles are 
	%simultaneously inverted 
	%after a preconditioner that corrects   
	%for the geometrical spreading effects.
}
The starting model for all inversion examples is
the same as the background homogeneous model  
with $\mVp = \textrm{2000}\,$m\,s$^{-1}$ and $\mR = \textrm{2000}\,$kg\,m$^{-3}$.
The peak frequency of the Ricker source wavelet, which is assumed to be known during inversion, is 5$\,$Hz.
The dominant wavelength of the acoustic wavefield used for imaging is 400$\,$m.

\begin{figure}
\begin{center}
	\includegraphics[width=\textwidth]{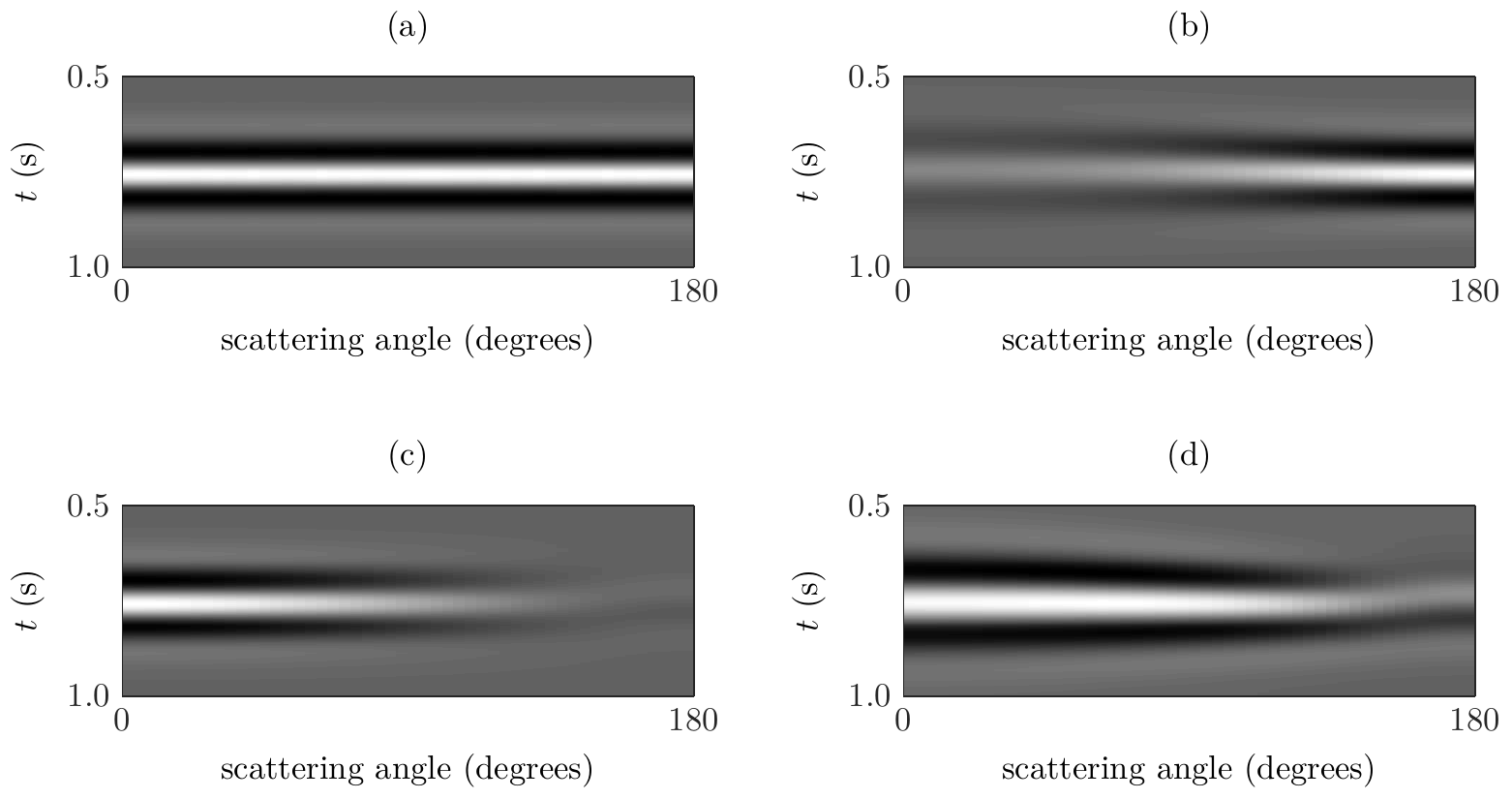}
%\begin{tikzpicture}
%
%\blockmatrix{6}{6}{he}{black}{white}{true}{black}{blue}{0.5}
%
%\end{tikzpicture}
	\caption{
		\revtext{Observed data 
	as a function of scattering angle due to: a) point $\mVp$-only scatterer (i); b) Gaussian-shaped $\mVp$-only scatterer (ii); 
	c)  point $\mR$-only scatterer (iv); d) Gaussian-shaped $\mR$-only scatterer (v).
	The data due to the
	Gaussian-shaped scatterers are different from those of the point scatterers, even though they have a size less than a quarter of the dominant wavelength.
	The scattering amplitude-versus-angle for point scatterers in these plots 
	is determined by the 
	radiation patterns in Figure~\ref{fig:param_radiation}.
	}
	}
\label{fig:param_data_point_gauss}
\end{center}
\end{figure}

If one aims to reconstruct model perturbations of a size comparable to
the wavelength that corresponds to the dominant frequency in the data,
then the problem may be considered as almost well-posed even with
band-limited data.
The standard deviation of the extended Gaussian-shaped scatterers
is 50$\,$m, corresponding to a size %such that their size, when the two sides of the Gaussian are taken into account, is 
around one-fourth of the dominant wavelength.
\revtext{
This means that the observed extended-scatterer data, as shown in the Figure~\ref{fig:param_data_point_gauss},
mainly
differ from the point-scatterer data in scattering amplitudes, but not in kinematics.
\wrmrk{Are you sure? With synthetics, it will affect the kinematics as well. Perhaps just a little bit, but there is no noise to mask the difference. }
\prmrk{added mostly}
Note the 
difference in 
amplitude-versus-angle between 
Figures~\ref{fig:param_data_point_gauss}a and \ref{fig:param_data_point_gauss}b
or 
\ref{fig:param_data_point_gauss}c and \ref{fig:param_data_point_gauss}d,
suggesting that the radiation-pattern analysis, which only considers point scatterers, 
is not valid for the
extended Gaussian-shaped sub-wavelength scatterers.
}

\begin{table}
\caption{Properties of different scatterers located at $\xx_0$.  }
\label{tab:param_scat}
\begin{center}
\begin{tabular}{ | l | c | c|  c|r |}
\hline
Scatterer &Actual Properties & Shape & Properties \\ 
Case & (Shape \& Contrast) & Known? & to be Estimated\\ \hline
(i) & Point; $\mVp$-only  & Yes & Contrast  \\ \hline
(ii) & Gaussian-shaped; $\mVp$-only  & Yes& Contrast \\ \hline
(iii) & Gaussian-shaped; $\mVp$-only  & No &Shape \& Contrast  \\ \hline
(iv) & Point; $\mR$-only  & Yes & Contrast  \\ \hline
(v) & Gaussian-shaped; $\mR$-only  & Yes& Contrast \\ \hline
(vi) & Gaussian-shaped; $\mR$-only  & No &Shape \& Contrast  \\ \hline
(vii) & Gaussian-shaped; non-$\mB$  & No &Shape \& Contrast  \\ \hline
\end{tabular}
\end{center}
\end{table}

For the parameterization analysis, we position one out of the 
seven different scatterers, listed in Table~\ref{tab:param_scat}, at $\xx_0$.
We employ the  
three modelling and inversion schemes to
reconstruct 
the scatterer at $\xx_0$ with different subsurface 
parameterizations (Table~\ref{tab:param_param}).
During inversion, 
we examine the
relative convergence rate
of different parameterizations by
displaying the least-squares \revtext{data} misfit  as a function of the iteration count on a log-log scale.
We want to observe if 
\begin{itemize} 
\item the suggestions of the point-scatterer analysis and/or radiation pattern analysis can be validated at least by one of the three modelling and inversion schemes used;

\item the relative convergence rate of a particular parameterization choice
is similar among 
different inversion schemes;

\item the relative convergence rate of a particular 
parameterization choice  
depends on the actual contrast of the scatterer --- we have chosen scatterers (i)--(iii) with a $\mVp$-only contrast, (iv)--(vi) with a $\mR$-only contrast and 
scatterer (vii) such that there is no contrast in $\mB$;

\item the relative convergence rate of a particular 
parameterization choice
depends on the scatterer properties that are to be estimated ---
only the contrast of the scatterers (i), (ii), (iv) and (v) has to be
estimated, while both the contrast and shape are unknown for other scatterers;

\item the relative convergence rate of a particular
parameterization choice during reconstructing point-shaped scatterers is different when compared to 
Gaussian-shaped scatterers ---
scatterers (i) and (iv) are point-shaped while the others are Gaussian-shaped.

\end{itemize}

\subsection{Error-bowl Analysis}
Since there are only 
two unknown  variables while 
reconstructing scatterers (i), (ii), (iv) and (v) with a known shape (see Table~\ref{tab:param_scat}),
we can plot 
two-dimensional logarithmic contours of the least-squares \revtext{data} misfit, i.e., contours of $-\logten(\funcls)$,
as a function of the variables.
These contours or error bowls can be used
to assess
the observed relative convergence rate of each  parameterization choice.
Similar to the analysis using the Hessian matrix,
the relative convergence rate 
for each parameterization can also be assessed based on the 
(a) ellipticity of the misfit contours;
(b) angle between the error vector and one of the principle axis of 
the contours.
We call this an error-bowl analysis.
In this analysis,
a more elliptic or less circular 
error bowl means 
that it is more
sensitive 
to one parameter than the other.
The orientation angle 
of an error bowl determines the correlation between the two parameters.
An orientation angle of 
$0^\circ$ or $90^\circ$
means that the parameters
are least correlated while
an angle of 
$45^\circ$ or $135^\circ$
means that they are most correlated.
So the best convergence rate 
is for a circular error bowl 
and 
the worst convergence rate is for an elliptical 
bowl
oriented at either 
$45^\circ$ or $135^\circ$
angle.

We are analysing the same simplified inverse problem as that of the point-scatterer analysis, 
when the error bowls are plotted for the point-shaped scatterers (i) and (iv) after adopting scheme I.
Therefore, the error-bowl analysis is the same as the point-scatterer analysis in this case, because
the shape and orientation of the error bowls in Figure~\ref{fig:param_bowls_scat1_scheme1} 
are determined 
by the reduced Hessians.
For example, the error bowls for \lxxrho- or \ilxirh-parameterization, in Figure~\ref{fig:param_bowls_scat1_scheme1},
are circular due to the fact that the condition number of 
$\Hker_{\sfrac{1}{\mB},\sfrac{1}{\mR}}$ or  $\Hker_{\mB,\mR}$ is one. 
Furthermore,
the error-bowl analysis can be seen as an extension 
of the point-scatterer analysis, in which
different modelling and inversion schemes can be employed
along with arbitrarily shaped scatterers.
It has to be noted that it is 
%cumbersome 
impossible
to derive analytical expressions of the reduced Hessians in the case of
arbitrarily shaped scatterers, when full-waveform modelling is employed.
\wrmrk{Can you actually derive those at all? Or do you resort to numerical
techniques anyway? What is the purpose of the statement?}
\prmrk{I wanted to say that we have to get away from analytics if we want 
to do arbitrary shapes and error-bowl analysis is a good numerical way to do it.}
%
%In the case of the Gaussian-shaped scatterers,
%we want to observe if the suggestions of the error-bowl analysis
%are validated when both the shape and contrast of the scatterers are unknown.

\subsection{Point-shaped Scatterers}

\begin{figure}
\begin{center}
	\includegraphics[width=\textwidth]{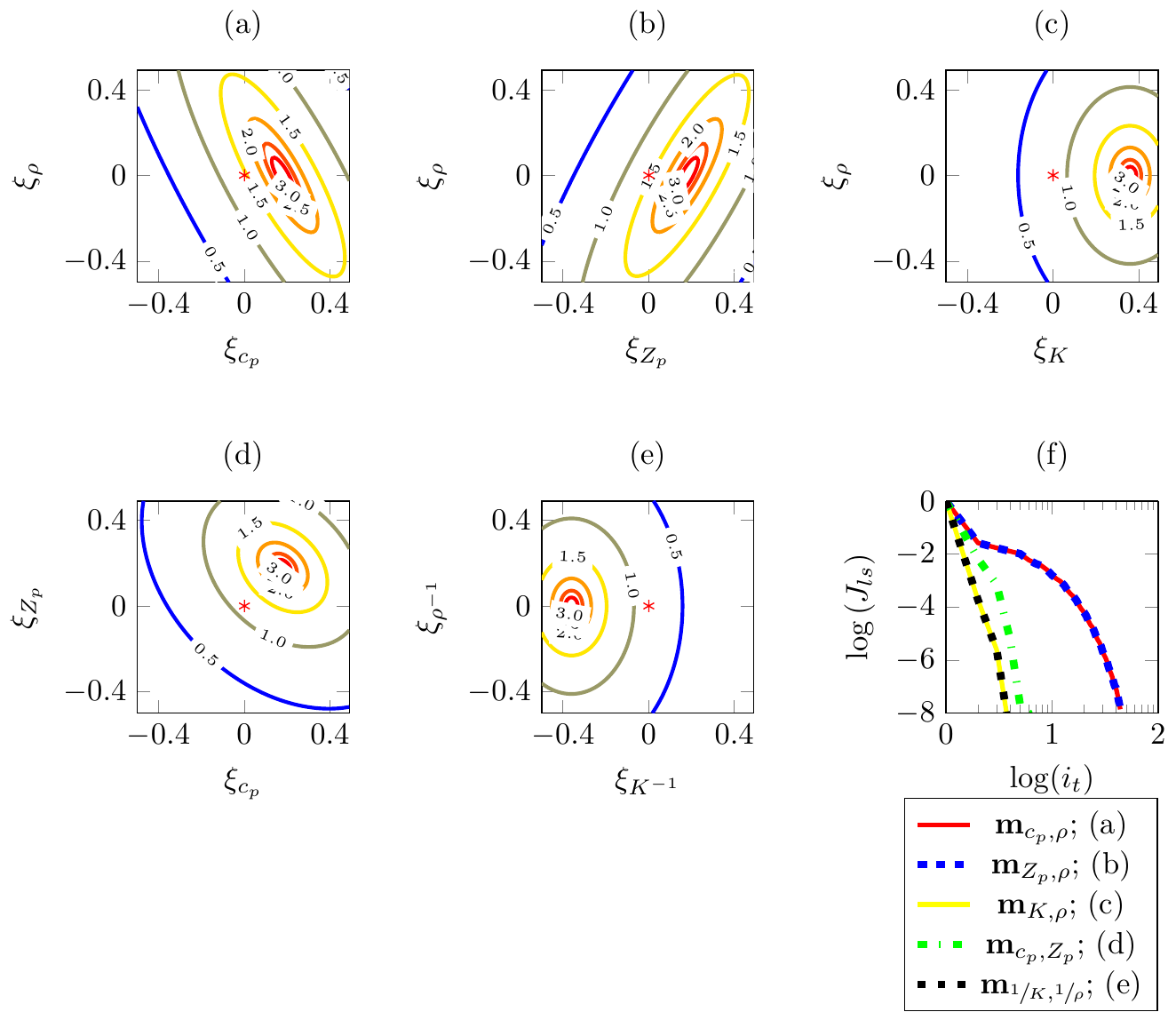}
\end{center}
\caption{
Reconstruction of the point-shaped $\mVp$-only scatterer (i) when the modelling and 
inversion scheme I, with the Born approximation and linear re-parameterization, is adopted.
	Error-bowl analysis is performed by plotting the logarithmic contours of the least-squares \revtext{data} misfit as a function of 
the two subsurface parameters while using the following parameterization choices: 
a) \velrho --- slower convergence expected due to high ellipticity;
b) \ipxrho --- slower convergence expected due to high ellipticity;
c) \lxxrho---expect faster convergence due to circular contours; 
d) \velipx---expect faster convergence due to almost circular contours;
e) \ilxirh---expect faster convergence due to circular contours.
In all the plots, the starting homogeneous model, $(0,0)^{\mathrm{T}}$, is marked by the red star.
	f) The least-squares \revtext{data} misfit is plotted as a function of the iteration count on a log-log scale. 
This plot shows that the 
suggestions of the point-scatterer analysis are valid in this case.
}
\label{fig:param_bowls_scat1_scheme1}
\end{figure}
\begin{figure}
\begin{center}
	\includegraphics[width=\textwidth]{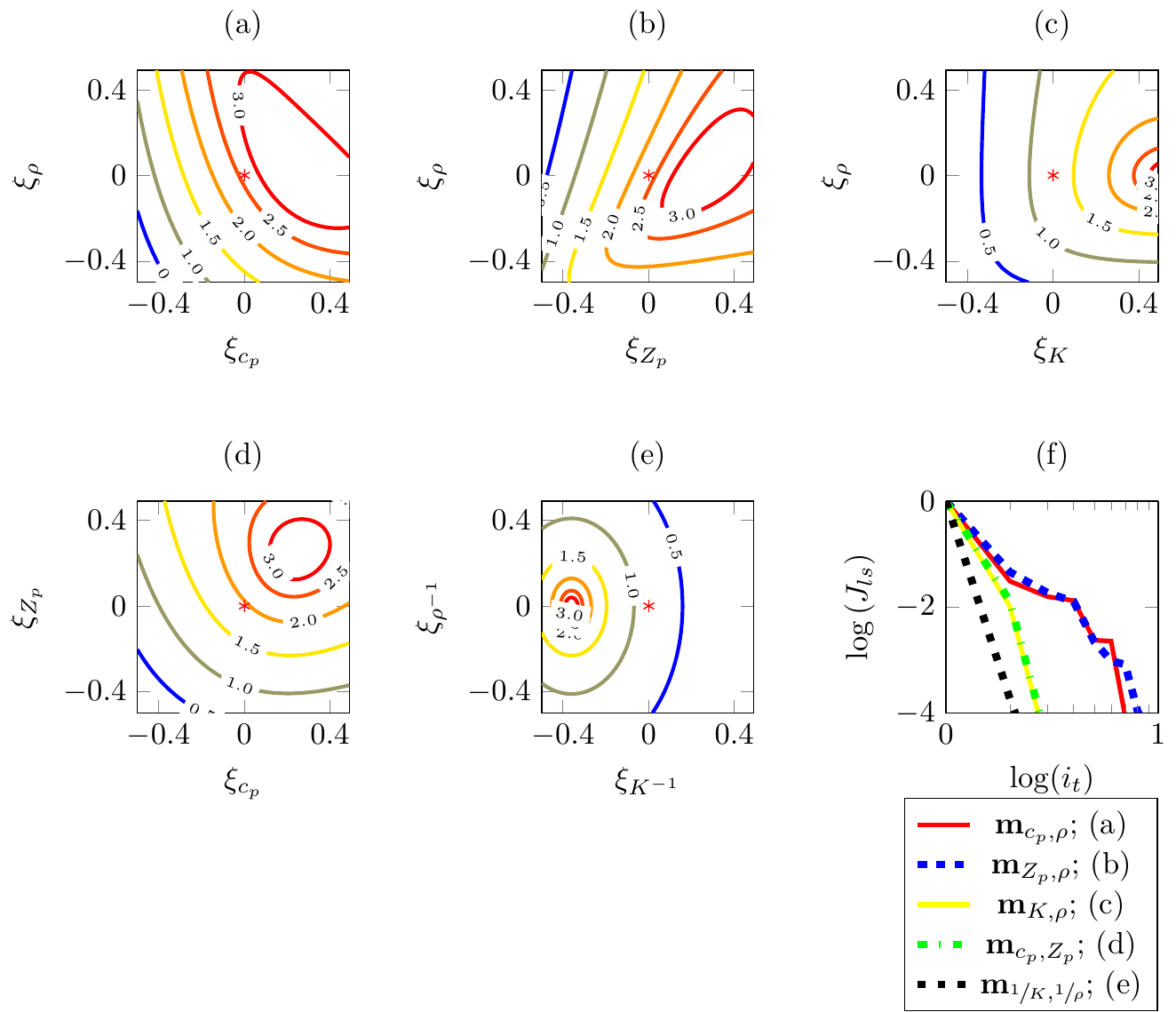}
\end{center}
\caption{
Same as Figure~\ref{fig:param_bowls_scat1_scheme1}, except for adopting the modelling and inversion scheme II, with the Born approximation and 
non-linear re-parameterization.
It can be seen 
that the error bowls for the \lxxrhop are not equivalent to the \ilxirhp 
because of the non-linear re-parameterization.
Also, the shapes of the error bowls (a)--(d) are different compared to that of Figure~\ref{fig:param_bowls_scat1_scheme1}.
It can be noted that 
the error bowls (c)--(e) 
are more \emph{circular} than error bowls (a)--(b).
The least-squares misfit plot shows that the 
suggestions of the point-scatterer analysis are valid.
}
\label{fig:param_bowls_scat1_scheme2}
\end{figure}
\begin{figure}
\begin{center}
	\includegraphics[width=\textwidth]{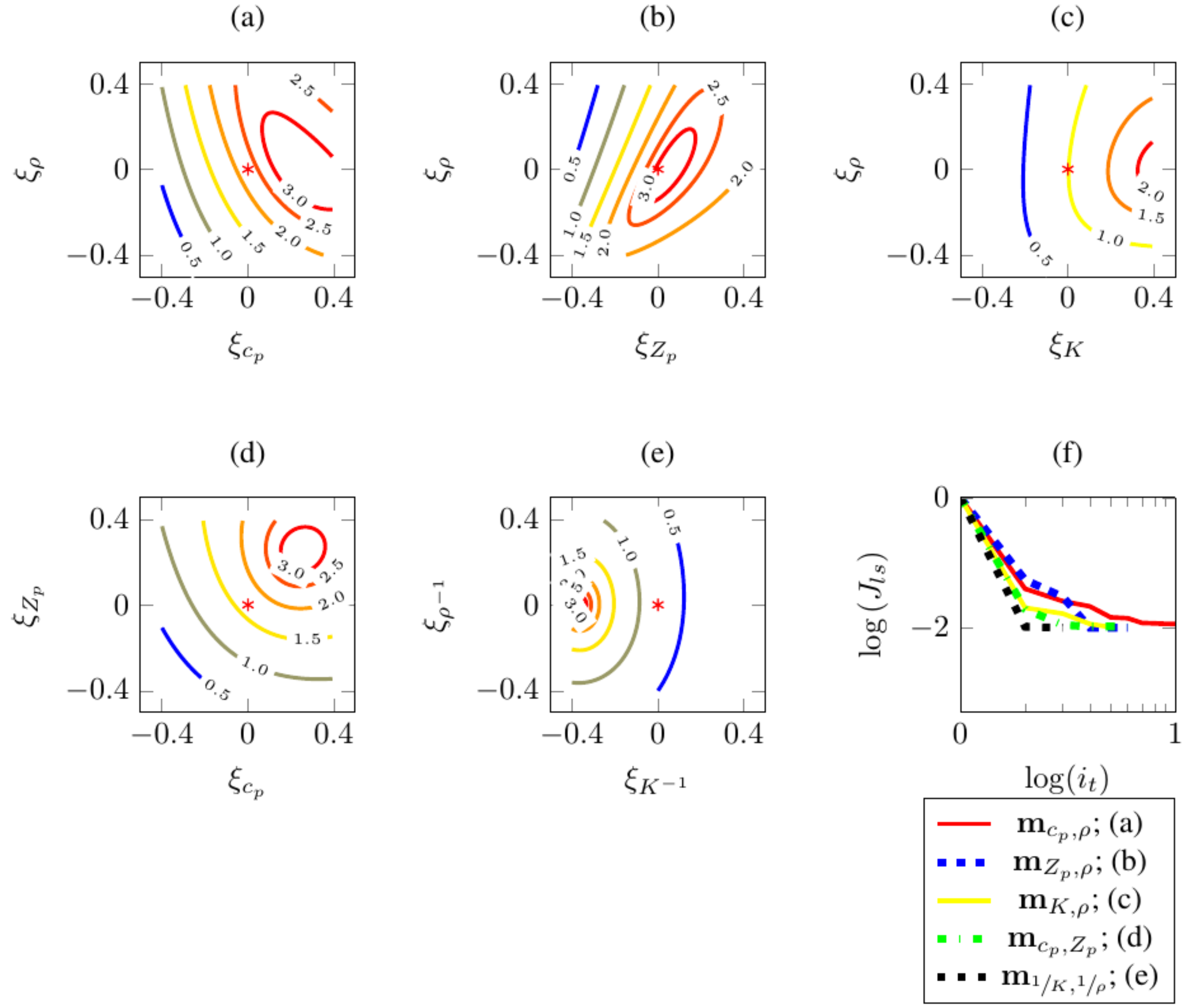}
\end{center}
\caption{
Same as Figure~\ref{fig:param_bowls_scat1_scheme1}, except for adopting scheme III, with full-waveform modelling and inversion.
We see that the 
shapes of the error bowls and the relative convergence rates of different parameterization choices
are similar to that of Figure~\ref{fig:param_bowls_scat1_scheme2}.
%
%\wrmrk{why no convergence?} \prmrk{I don't understand why, maybe it has to do with my finite-difference code using single-precision}
The least-squares misfit plot 
shows that 
the 
suggestions of the point-scatterer analysis are 
valid.
}
\label{fig:param_bowls_scat1_scheme3}
\end{figure}

We first consider the 
inverse problems of reconstructing
the point-shaped scatterer (i) with $\mVp$-only contrast in Table~\ref{tab:param_scat}.
Appendix A, which reconstructs an equivalent scatterer (iv) with $\mR$-only contrast, provides additional support to the discussion in this subsection.

When scheme I is adopted for modelling and inversion, 
the error-bowl analysis, identical to the point-scatterer analysis, suggests  
that \lxxrho- or \ilxirh-parameterizations 
have the fastest convergence 
because of their corresponding circular error bowls, as plotted in 
Figures~\ref{fig:param_bowls_scat1_scheme1}c and \ref{fig:param_bowls_scat1_scheme1}e.
In addition to that, we observe that 
the bowls for  \velipxp are more circular than either the
\velrho- or \ipxrho-parameterization.
The orientation also explains the slightly slower convergence.
The least-squares misfit plotted against the iteration count 
while estimating the $\mVp$-only contrast, in Figure~\ref{fig:param_bowls_scat1_scheme1}f 
shows that the \velipx-, \lxxrho- and \ilxirh-parameterizations have 
faster convergence than the others. 

We now employ the modelling and inversion scheme II.
Figure~\ref{fig:param_bowls_scat1_scheme2} shows that the 
ellipticity of the error bowls for all the parameterizations, except for the \ilxirhp, are different from the previous case.
Note that 
the \lxxrhop 
is no longer equivalent to the
\ilxirhp because of the non-linear re-parameterization. 
However, similar to the previous case with the scheme I, we observe that the 
\velipx-, \lxxrho- and \ilxirh-parameterizations converge faster.

Figure~\ref{fig:param_bowls_scat1_scheme3} 
display the error bowls when modelling and inversion with 
scheme III is employed.
It can be seen that the suggestions 
of the point-scatterer analysis
remain valid even when employing the modelling and inversion scheme II or III.
This is because the error bowls (c)--(e)
are more \emph{circular} or less \emph{elliptical} and more oriented towards 
$0^\circ$ or $90^\circ$
than the error bowls (a)--(b) in the 
Figures~\ref{fig:param_bowls_scat1_scheme2} and
\ref{fig:param_bowls_scat1_scheme3}.

\subsection{Gaussian-shaped Scatterers}

\begin{figure}
\begin{center}
	\includegraphics[width=\textwidth]{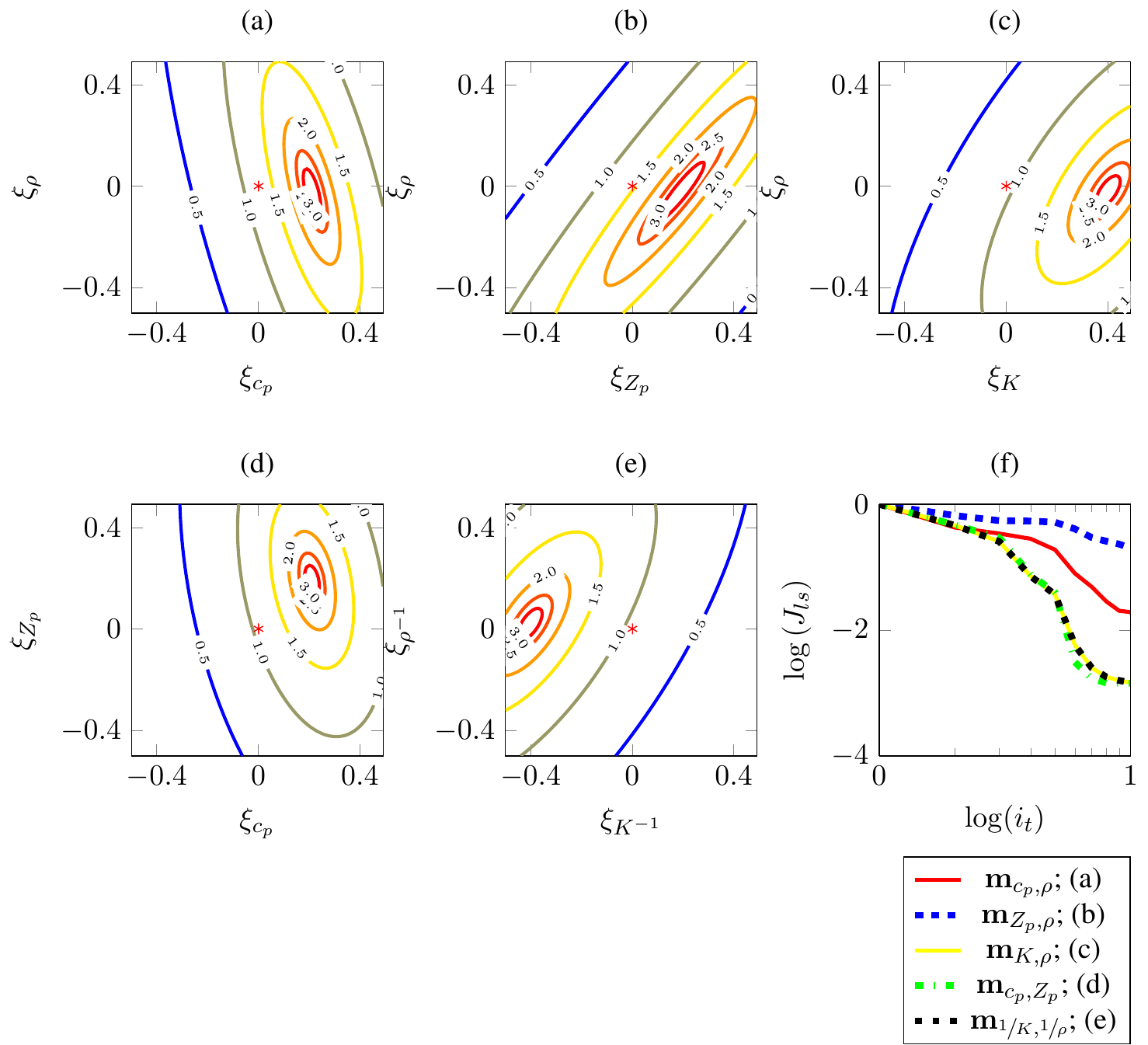}
\end{center}
\caption{
Same as Figure~\ref{fig:param_bowls_scat1_scheme1}, except for the Gaussian-shaped $\mVp$-only scatterer (ii).
We observe that the errors bowls for all the parameterization choices are more elliptical compared to that of Figure~\ref{fig:param_bowls_scat1_scheme1}.
The least-squares misfit plot shows that the 
suggestion of the 
point-scatterer analysis is valid. 
The contours in the case \ilxirh- and \lxxrho-parameterizations are not circular.
The rate of convergence in the case of \velipx-, \lxxrho- and \ilxirh-parameterizations 
is higher as their corresponding error bowls are more circular compared to the others.
}
\label{fig:param_bowls_scat2_scheme1}
\end{figure}
\begin{figure}
\begin{center}
	\includegraphics[width=\textwidth]{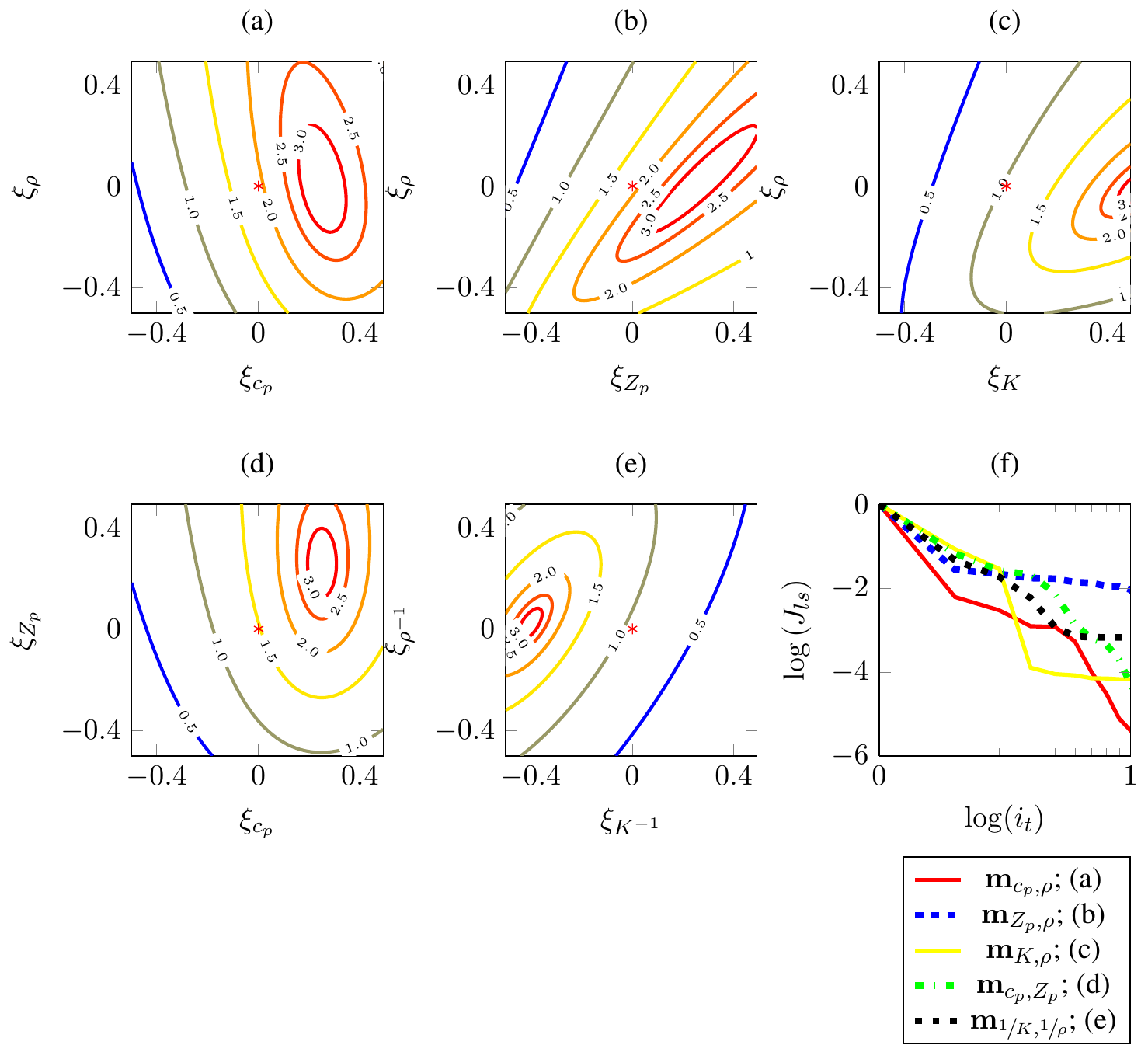}
\end{center}
\caption{
Same as Figure~\ref{fig:param_bowls_scat1_scheme1}, except for scatterer (ii) and adopting scheme II, with the Born approximation and non-linear re-parameterization.
Compared to the Figure~\ref{fig:param_bowls_scat2_scheme1}, we see that 
the error bowls, in this case, have much different orientation.
The least-squares misfit plot shows that the 
suggestions of the 
point-scatterer analysis are no longer valid. 
}
\label{fig:param_bowls_scat2_scheme2}
\end{figure}
\begin{figure}
\begin{center}
\includegraphics[width=\textwidth]{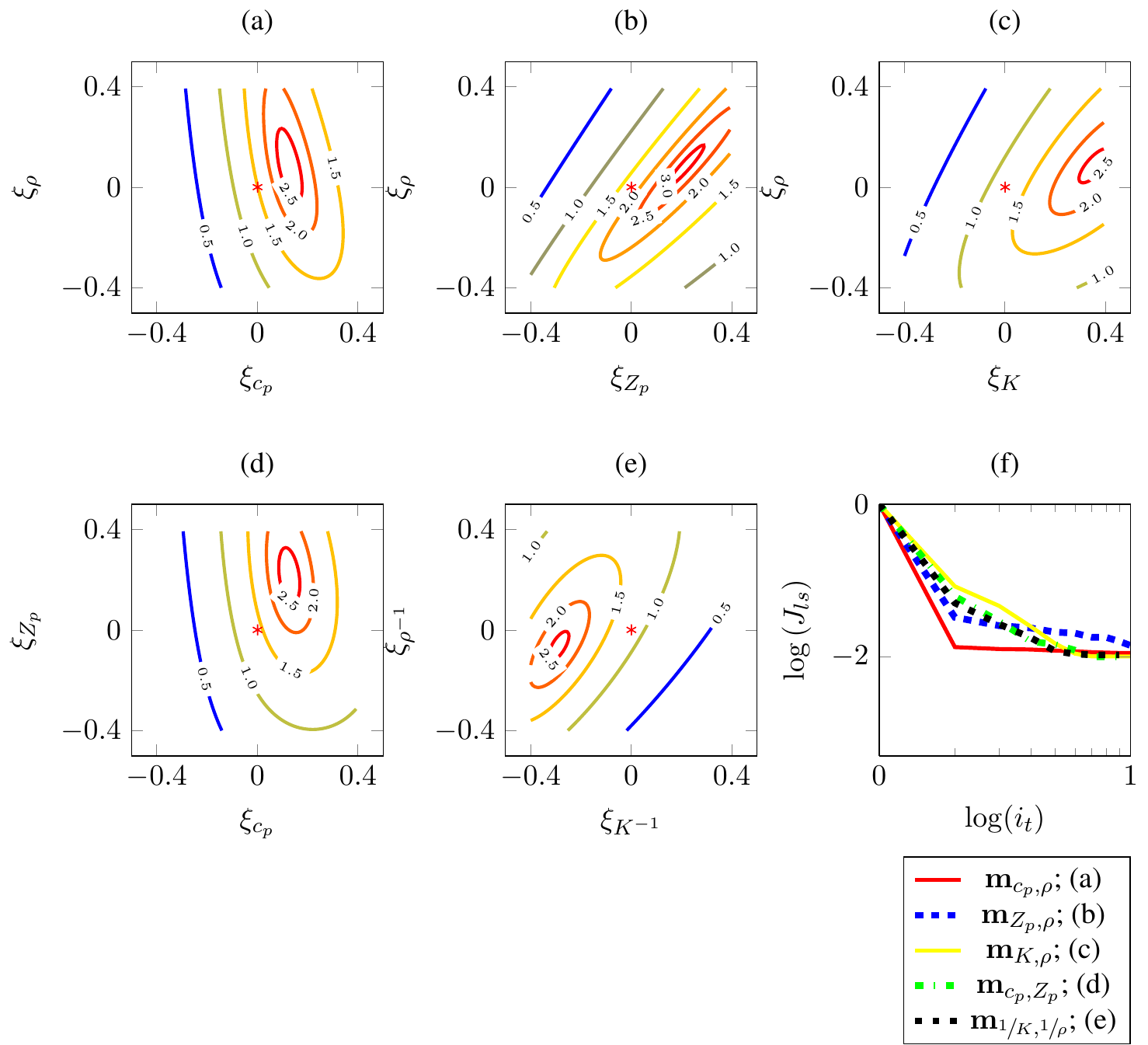}
\end{center}
\caption{
Same as Figure~\ref{fig:param_bowls_scat1_scheme1}, except for scatterer (ii) and adopting scheme III, with full-waveform modelling and inversion.
The least-squares misfit plot shows that the 
suggestions of the 
point-scatterer analysis are no longer valid. 
Parameterization using \velrho has the best convergence rate. 
}
\label{fig:param_bowls_scat2_scheme3}
\end{figure}

We now want to see
if the sub-wavelength Gaussian-shaped scatterer (ii) with $\mVp$-only contrast at $\xx_0$ will 
result in error bowls of a different shape  
compared to the point-shaped scatterers of the previous subsection.
Appendix A, which reconstructs an equivalent scatterer (v) with $\mR$-only contrast, 
provides additional support to the discussion in this subsection.
When the modelling and inversion scheme I is employed,
in Figure~\ref{fig:param_bowls_scat2_scheme1}, 
the error bowls are more elliptical for the \velrho-, \ipxrho- and \velipx-parameterization choices
than those with point-shaped scatterers in Figure~\ref{fig:param_bowls_scat1_scheme1}.
The bowl of \ipxrhop is maximally oriented towards a 
$45^\circ$
angle compared to the others.
Moreover, the error bowls for the  \lxxrho- and \ilxirh-parameterizations are not 
circular anymore.
We still observe that 
the \velipx-, \lxxrho- and \ilxirh-parameterizations have a 
faster convergence  because their corresponding error bowls are more circular.

When either modelling and inversion scheme II or III is employed, 
the error bowls for all parameterizations not only have different ellipticities 
but also different orientations.
This indicates that the suggestions of the point-scatterer analysis are no
longer valid.
We observe that the \velrhop is faster in reconstructing a $\mVp$-only 
contrast (Figure~\ref{fig:param_bowls_scat2_scheme2}f and \ref{fig:param_bowls_scat2_scheme3}f) 
and that the \ilxirhp has the fastest convergence when
reconstructing a $\mR$-only contrast (see Appendix). 
Hence, when modelling and inversion schemes II and III are employed to reconstruct non-point shaped scatterers,
the rate of convergence of a particular parameterization also depends on 
the type of contrast that has to be reconstructed.

\subsection{Scatterers with Unknown Shape and Contrast}

\begin{figure}
\begin{center}
\includegraphics[width=\textwidth]{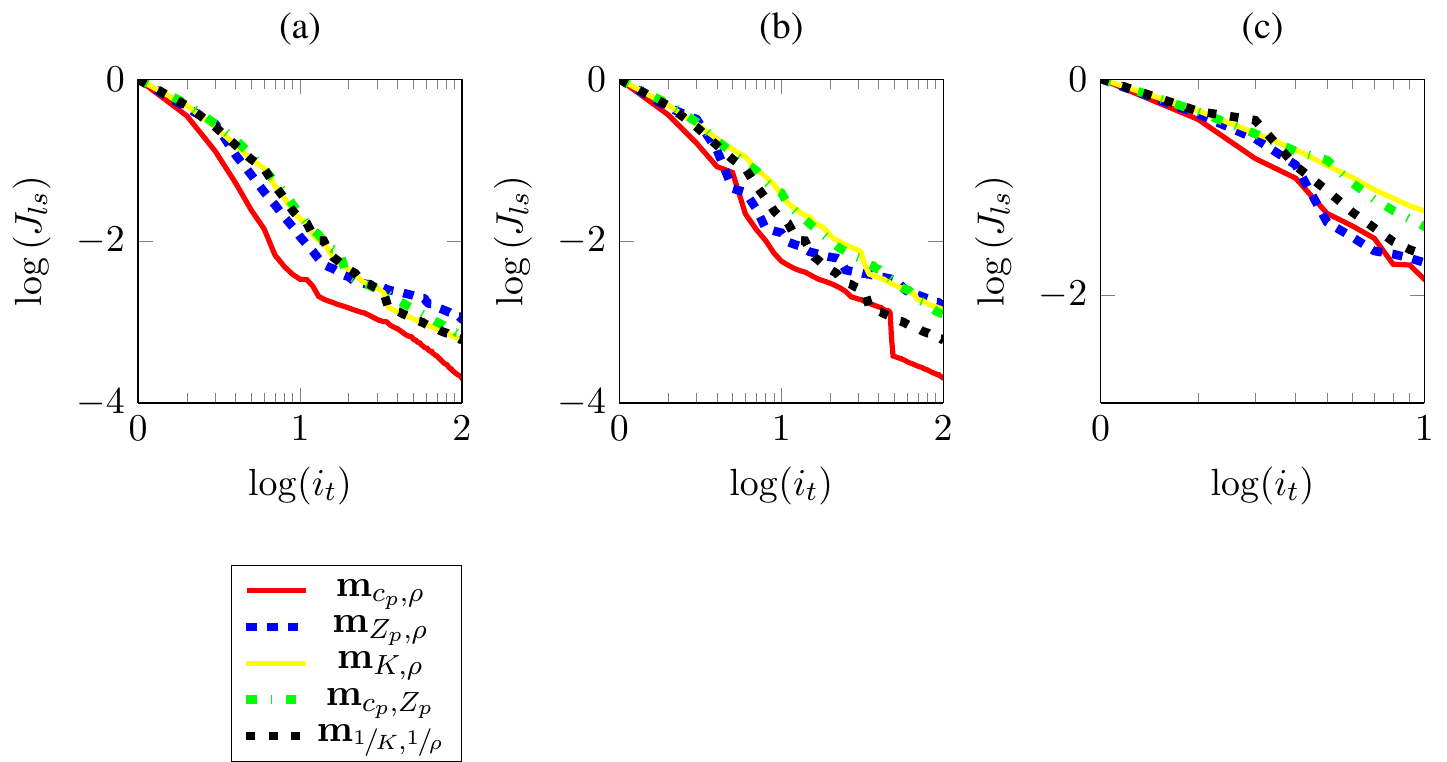}
\end{center}
\caption{
The least-squares misfit is plotted as a function of the iteration count on a log-log scale during the 
reconstruction of scatterer (iii), where both shape and contrast of a $\mVp$-only scatterer are unknown.
The modelling and inversion schemes (a) I, (b) II and (c) III are adopted. 
It can be observed that the parameterization using \lxxrho has relatively lower convergence rate when all the three cases are taken into account.
}
\label{fig:param_bowls_scat3_all}
\end{figure}

\begin{figure}
\begin{center}
	\includegraphics[width=\textwidth]{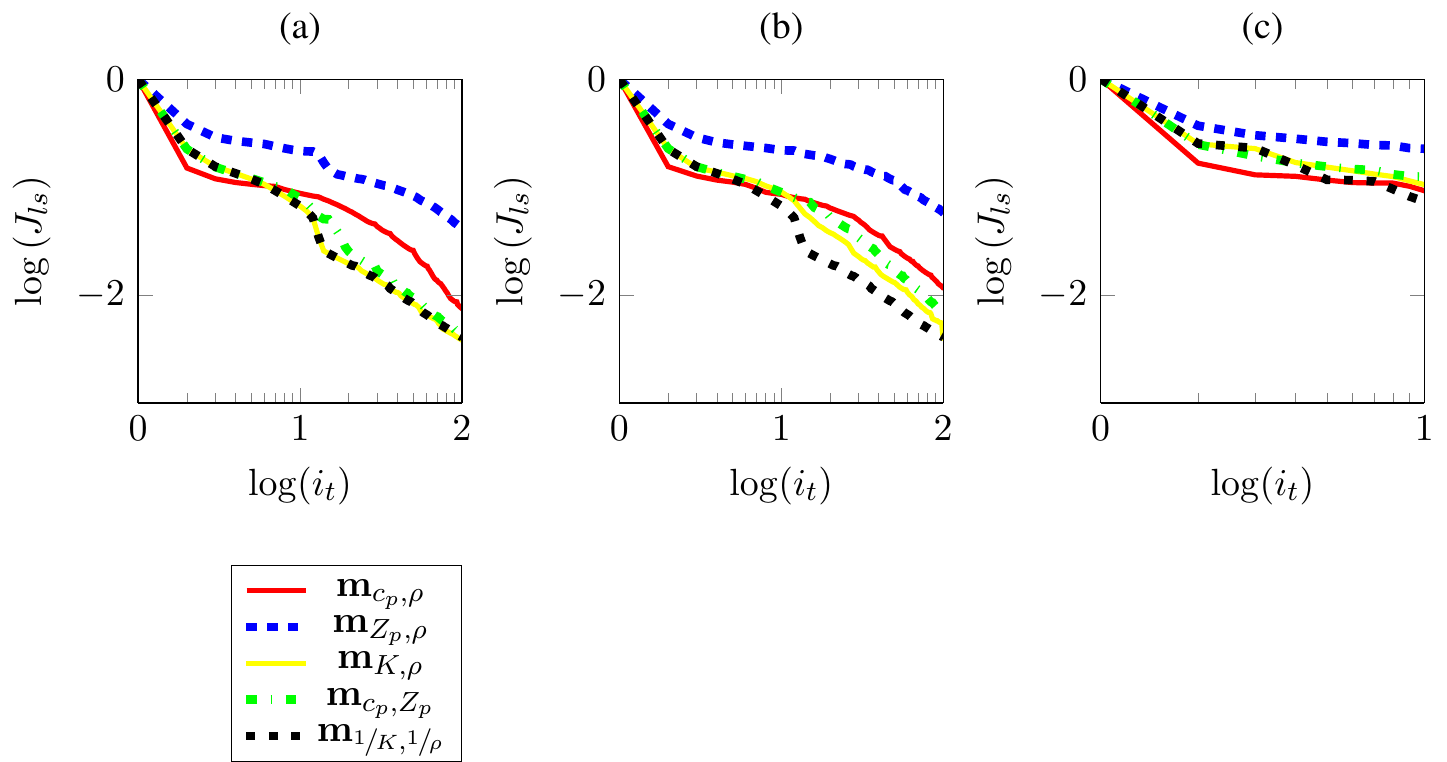}
\end{center}
\caption{
Same as Figure~\ref{fig:param_bowls_scat3_all}, except for scatterer (vi), with $\mR$-only contrast.
Note that the
relative rates of convergence of various parameterization choices are different compared to that of 
the Figure~\ref{fig:param_bowls_scat3_all}.
It can be observed that the parameterization using \ipxrho has the worst convergence rate in all the cases.
}
\label{fig:param_bowls_scat6_all}
\end{figure}

\begin{figure}
\begin{center}
	\includegraphics[width=\textwidth]{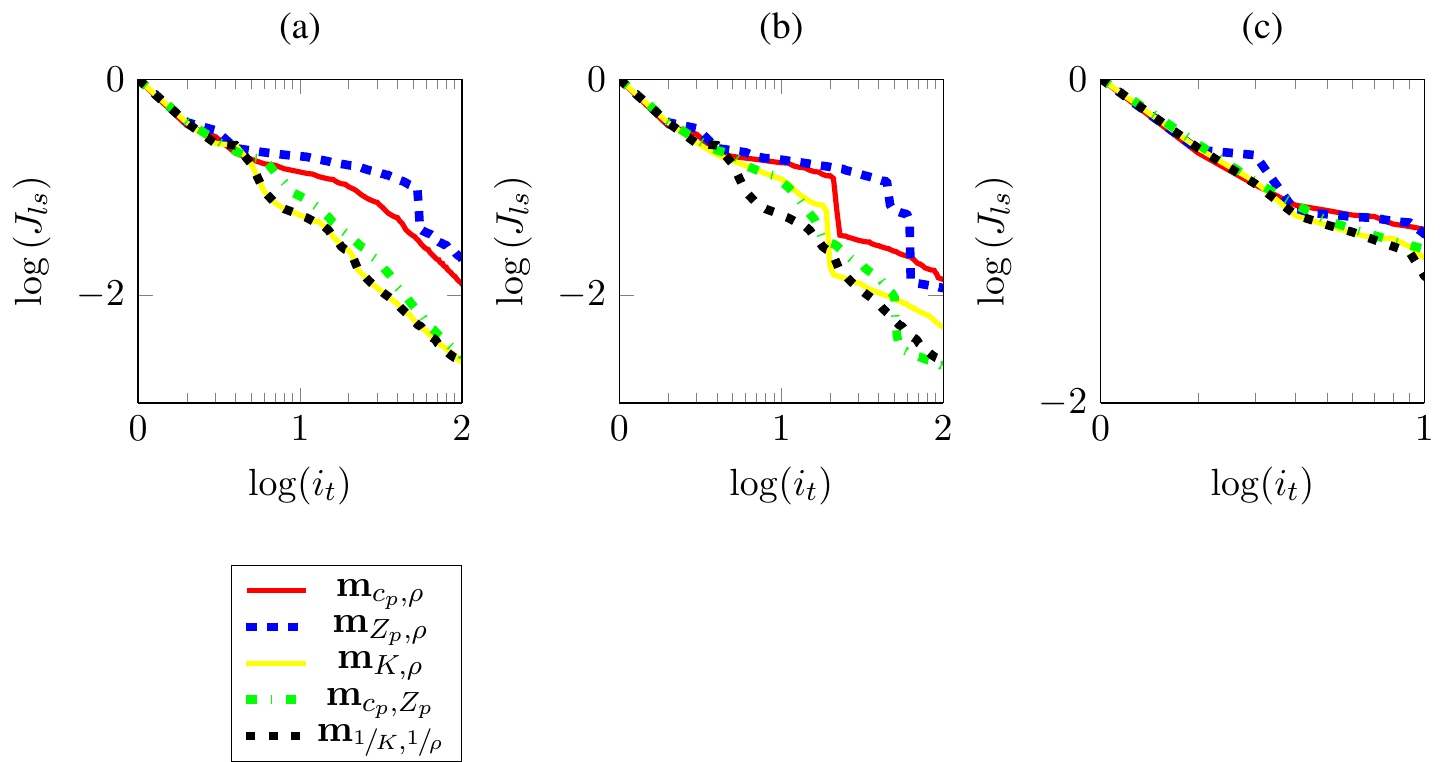}
\end{center}
\caption{
Same as Figure~\ref{fig:param_bowls_scat3_all}, except for scatterer (vii), with a non-$\mB$ contrast.
Note that relative rates of convergence of various parameterization choices are different compared to that of 
the Figures~\ref{fig:param_bowls_scat3_all} and \ref{fig:param_bowls_scat6_all}.
It can be observed that the parameterization using \ipxrho has the worst convergence rate in all the cases.
}
\label{fig:param_bowls_scat7_all}
\end{figure}

We now consider the inverse problems of estimating both the shape and contrast of the sub-wavelength 
Gaussian scatterers (iii), (vi) and (vii).
The error bowls cannot be easily plotted in this case
because there are more than two unknown variables.
As shown in Figures~\ref{fig:param_bowls_scat3_all}, 
\ref{fig:param_bowls_scat6_all} and \ref{fig:param_bowls_scat7_all},
the \velrhop is the fastest 
while estimating a $\mVp$-only scatterer,
whereas the \velipx-, \lxxrho- and \ilxirh-parameterizations have a faster convergence 
while 
estimating a $\mR$-only or a non-$\mB$ scatterer. 
This observation is similar to that of estimating the contrast of the 
Gaussian-shaped scatterers in the last subsection. 

\section{Hierarchical Inversion}
\label{sec:nonuni}
\begin{figure}
\begin{center}
\includegraphics[width=\textwidth]{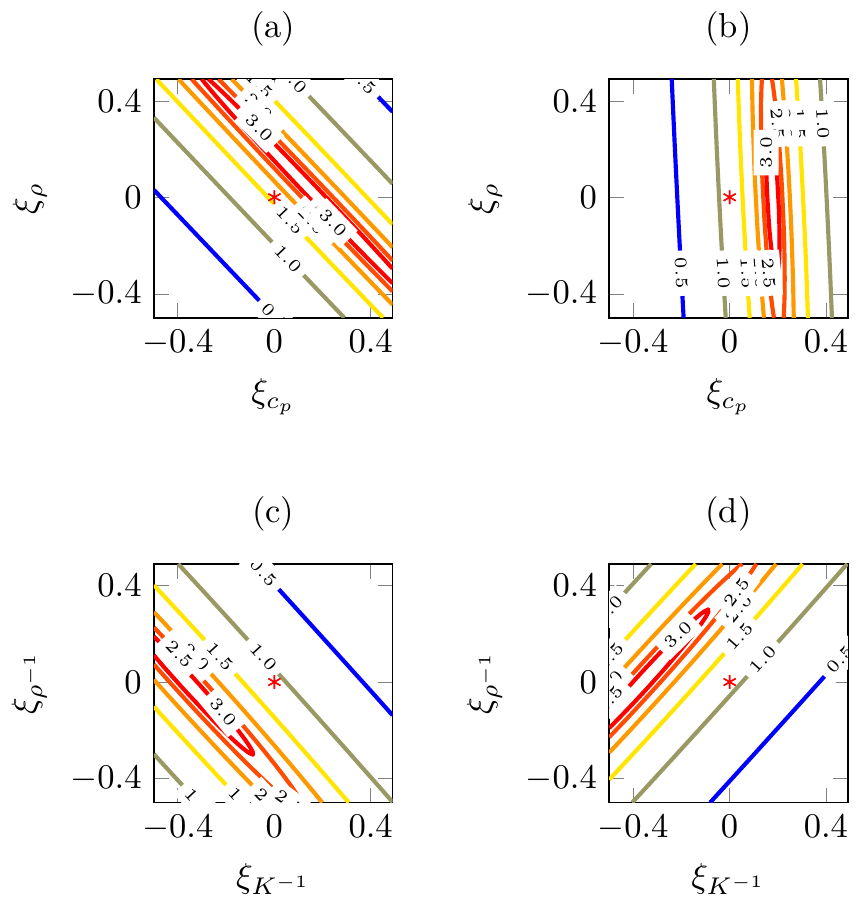}
\end{center}
\caption{
	\revtext{
Similar to Figure~\ref{fig:param_bowls_scat1_scheme1}, except 
	a) using \velrhop and short-offset regime with $0^\circ$--$90^\circ$ scattering angles only;
	b) using \velrhop and long-offset regime with $135^\circ$--$180^\circ$ scattering angles only;
	c) same as (a), but for \ilxirhp;
	d) same as (b), but for \ilxirhp.
	}
}
\label{fig:non_unique}
\end{figure}

\revtext{
	Let us
	consider 
	the problem of finding the best parameterization choice for a multi-parameter
	inversion with limited-aperture acquisition.
	Such inversions are common during 
	hierarchical data-fitting approaches, for example, from wide angles to short offsets.
In Figure~\ref{fig:non_unique},
we plotted 
the \velrho and \ilxirh 
error bowls similar to Figure~\ref{fig:param_bowls_scat1_scheme1}, except in
two separate regimes:
\begin{itemize}
	\item short offsets with
		$0^\circ$--$90^\circ$ scattering angles only;
	\item long offsets with
		$135^\circ$--$180^\circ$ scattering angles only.
\end{itemize}
In both the regimes, it can be seen that 
the lack of angle-dependent 
information has resulted in 
error bowls with long \emph{valleys}
pointing in the \emph{null-space} direction.
Lack of angle-dependent
information in the data makes it difficult to converge to the true solution.
%
% Which means,
Depending on the 
starting model, the inversion just chooses 
the shortest possible path to reach the valley and stagnates.
% why the well-posed problem?
%
The acoustic inverse problem 
is non-unique when the source-receiver aperture is limited and/or 
the necessary frequencies are lacking from the data.
%
%If the non-uniqueness due to the absence of high frequencies is
%ignored, there is always the imprint of a finite acquisition geometry.
%
It is known that 
in the presence of non-uniqueness, different choices
of parameterization may lead to different 
inversion results (see, for example, \cite{bharadwaj2014param}), each
explaining the data at convergence.
Hence, the search for the \emph{best} parameterization in terms of
convergence speed might be obfuscated by non-uniqueness problems 
for multi-parameter inversion
in the separated regimes that are
described above.
%
%
%In other words, the model obtained after 
%inversion
%depends on the 
%choice of parameterization.
%
%Therefore, due to these issues, 
%it is meaningless to perform 
%parameterization analysis for multi-parameter inversion 
%
However, 
one can still perform mono-parameter
inversion
in the 
subspace 
orthogonal 
to the null-space, 
by assuming a linearized problem,
in each of these regimes.
}

\section{Interpretation}
We now interpret 
the least-squares misfit graphs
in Figures~\ref{fig:param_bowls_scat1_scheme1}--\ref{fig:param_bowls_scat7_all} and \ref{fig:param_bowls_scat4_scheme1}--\ref{fig:param_bowls_scat5_scheme3},
which are plotted as a function of the iteration count  
during the reconstruction of
different scatterers in Table~\ref{tab:param_scat}.
In Table~\ref{tab:param_best},
the 
parameterization choices are divided 
into three classes depending on the relative convergence rate:
the choices with the faster and slower convergence 
are coloured blue and red, respectively;
the choices which neither have a faster nor a slower convergence 
are coloured in black. 
Table~\ref{tab:param_best} also indicates
if there is a significant difference in the convergence rate  among
the parameterizations.

We observe that there is
no best parameterization in a general sense since parameterizations
(C), (D) and (E) perform 
similarly good in most of the cases.
Also, note that these parameterizations have a slower convergence when 
reconstructing scatterer (iii).
However, on the other side of the spectrum, we observe that the \ipxrhop (case(B)) 
has the worst convergence rate, because it is 
coloured red or black almost in all the cases of Table~\ref{tab:param_best}. 
It can also be noticed that the \velrhop (case (A)) 
is the second worst one, because there is no significant difference in its convergence
even when it is the fastest.

\begin{table}
\caption{Reconstruction of seven different scatterers in Table~\ref{tab:param_scat} when different modelling and inversion
schemes are employed. The parameterizations, listed in 
Table~\ref{tab:param_param}, with 
faster and slower convergence relative to each other are in blue and red, respectively.
}
\label{tab:param_best}
\begin{center}
\begin{tabular}{ | l | c | c| c| r |}
\hline
Scatterer & Scheme I & Scheme II & Scheme III\\ \hline
(i) &\otext{(A)}~\otext{(B)}~\gtext{(C)}~\gtext{(D)}~\gtext{(E)} & \otext{(A)}~\otext{(B)}~\gtext{(C)}~\gtext{(D)}~\gtext{(E)}& \otext{(A)}~\otext{(B)}~\gtext{(C)}~\gtext{(D)}~\gtext{(E)}  \\ & Figure~\ref{fig:param_bowls_scat1_scheme1} & Figure~\ref{fig:param_bowls_scat1_scheme2} & Figure~\ref{fig:param_bowls_scat1_scheme3} \\ \hline
(ii) & \btext{(A)}~\otext{(B)}~\gtext{(C)}~\gtext{(D)}~\gtext{(E)}& 
\btext{(A)}~\otext{(B)}~\btext{(C)}~\btext{(D)}~\btext{(E)}& 
\gtext{(A)}~\btext{(B)}~\btext{(C)}~\btext{(D)}~\btext{(E)} \\ 
& &  & (no significant difference)\\ 
& Figure~\ref{fig:param_bowls_scat2_scheme1} & Figure~\ref{fig:param_bowls_scat2_scheme2} & Figure~\ref{fig:param_bowls_scat2_scheme3} \\\hline
(iii) & \gtext{(A)}~\btext{(B)}~\btext{(C)}~\btext{(D)}~\btext{(E)} & 
\gtext{(A)}~\btext{(B)}~\btext{(C)}~\btext{(D)}~\btext{(E)} & 
\btext{(A)}~\btext{(B)}~\otext{(C)}~\otext{(D)}~\btext{(E)} \\ 
& & (no significant difference) & (no significant difference)\\ 
& Figure~\ref{fig:param_bowls_scat3_all}a & Figure~\ref{fig:param_bowls_scat3_all}b & Figure~\ref{fig:param_bowls_scat3_all}c \\\hline
(iv) &\otext{(A)}~\otext{(B)}~\gtext{(C)}~\gtext{(D)}~\gtext{(E)} & \otext{(A)}~\otext{(B)}~\gtext{(C)}~\gtext{(D)}~\gtext{(E)} & \otext{(A)}~\otext{(B)}~\gtext{(C)}~\gtext{(D)}~\gtext{(E)} \\ & Figure~\ref{fig:param_bowls_scat4_scheme1} & Figure~\ref{fig:param_bowls_scat4_scheme2} & Figure~\ref{fig:param_bowls_scat4_scheme3} \\\hline
(v) &\btext{(A)}~\otext{(B)}~\gtext{(C)}~\gtext{(D)}~\gtext{(E)} &
\btext{(A)}~\otext{(B)}~\gtext{(C)}~\gtext{(D)}~\gtext{(E)} & 
 \btext{(A)}~\btext{(B)}~\btext{(C)}~\btext{(D)}~\gtext{(E)}\\ & 
Figure~\ref{fig:param_bowls_scat5_scheme1} & Figure~\ref{fig:param_bowls_scat5_scheme2} & Figure~\ref{fig:param_bowls_scat5_scheme3} \\\hline
(vi) &\gtext{(A)}~\otext{(B)}~\gtext{(C)}~\gtext{(D)}~\gtext{(E)} &
\gtext{(A)}~\otext{(B)}~\gtext{(C)}~\gtext{(D)}~\gtext{(E)} & 
\gtext{(A)}~\otext{(B)}~\gtext{(C)}~\gtext{(D)}~\gtext{(E)}  \\ 
& & & (no significant difference)\\ 
& Figure~\ref{fig:param_bowls_scat6_all}a & Figure~\ref{fig:param_bowls_scat6_all}b & Figure~\ref{fig:param_bowls_scat6_all}c \\\hline
(vii) &\otext{(A)}~\otext{(B)}~\gtext{(C)}~\gtext{(D)}~\gtext{(E)} &
\otext{(A)}~\otext{(B)}~\gtext{(C)}~\gtext{(D)}~\gtext{(E)} & 
\btext{(A)}~\btext{(B)}~\btext{(C)}~\btext{(D)}~\btext{(E)}  \\ 
& & & (no significant difference)\\ 
& Figure~\ref{fig:param_bowls_scat7_all}a & Figure~\ref{fig:param_bowls_scat7_all}b & Figure~\ref{fig:param_bowls_scat7_all}c \\ \hline
\end{tabular}
\end{center}
\end{table}

\section{Conclusions}
We have briefly outlined 
the conventional point-scatterer and the diffraction-pattern parameterization-analysis methods for the 
2-D acoustic inverse problem.
Using almost well-posed-numerical examples, we have shown that the suggestions 
of these conventional methods are valid 
only when the contrasts of the point-shaped scatterers 
at a known location are estimated.
The numerical examples employ three different 
modelling and inversion schemes
using both Born and full-waveform modelling.
As expected, for almost well-posed inverse problems, we observed that a change in 
parameterization will result in a different convergence rate.
Furthermore, the relative rate of convergence 
for a particular choice of parameterization depends on
(a) the modelling and inversion scheme employed;
(b) the contrast of  
subsurface scatterer that has to be reconstructed;
(c) the shape of the sub-wavelength scatterer.
We observed by considering all the cases that the
\ipxrhop 
has the worst convergence rate, followed 
by the \velrhop.
Finally, our numerical examples show that, in general, there 
is no such thing as the best parameterization choice that provides the 
fastest convergence for acoustic inversion.

\section{Acknowledgments}
The first author thanks Prof. Kees Wapenaar (TU
Delft) for helpful discussions. 
This work is a part of the NeTTUN project, funded from
the European Commissions Seventh Framework Programme for Research,
Technological Development and Demonstration (FP7 $2007$-$2013$) under
Grant Agreement $280712$.
The use of supercomputer facilities
was sponsored by % NWO Exacte Wetenschappen (Physical Sciences) with financial support from 
the `Nederlandse Organisatie voor Wetenschappelijk Onderzoek'
(Netherlands Organisation for Scientific Research, NWO).
The computations were carried out on the Dutch national 
e-infrastructure maintained by the SURF Foundation (www.surfsara.nl).

%\begin{thebibliography} 

%\end{thebibliography}

\bibliographystyle{acm} 
\bibliography{ch_param.bib}
%\references{./ch_param/ch_param.bib}

\clearpage
\appendix
\section{Error-bowls for Scatterers (iv) and (v)}
\begin{figure}
\begin{center}
	\includegraphics[width=\textwidth]{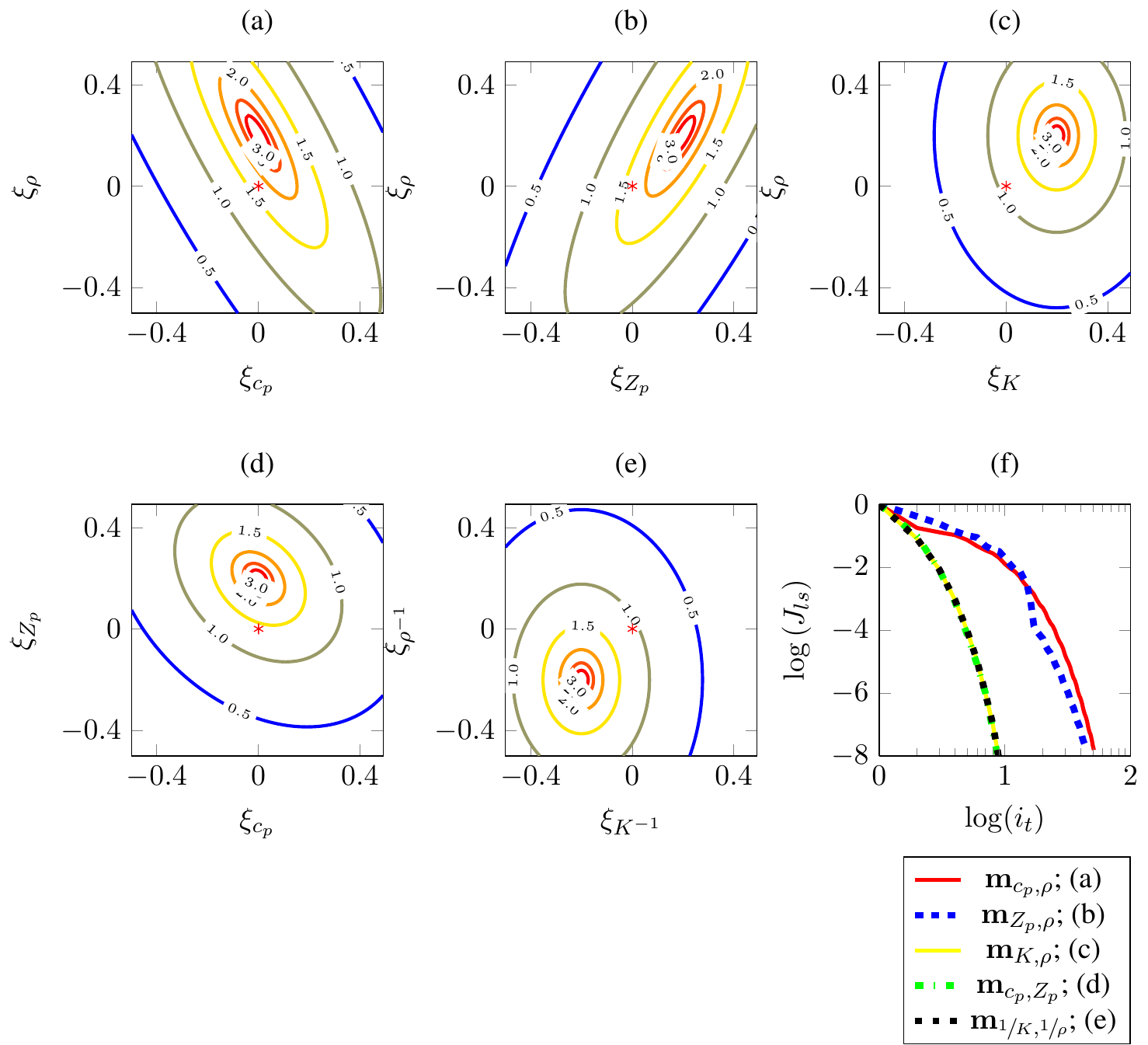}
\end{center}
\caption{
Reconstruction of scatterer (iv) using scheme I and the following parameterization choices:
a) \velrho --- slower convergence expected because of high ellipticity;
b) \ipxrho --- slower convergence expected because of high ellipticity;
c) \lxxrho --- faster convergence expected because of circular contours; 
d) \velipx --- faster convergence expected because of almost circular contours;
e) \ilxirh --- faster convergence expected because of circular contours.
In all the plots, the starting homogeneous model, $(0,0)^{\mathrm{T}}$, is marked by the red star.
f) The least-squares misfit is plotted as a function of the iteration count on a log-log scale. 
This plot shows that the 
suggestions of the point-scatterer analysis are valid in this case.
}
\label{fig:param_bowls_scat4_scheme1}
\end{figure}

\begin{figure}
\begin{center}
\includegraphics[width=\textwidth]{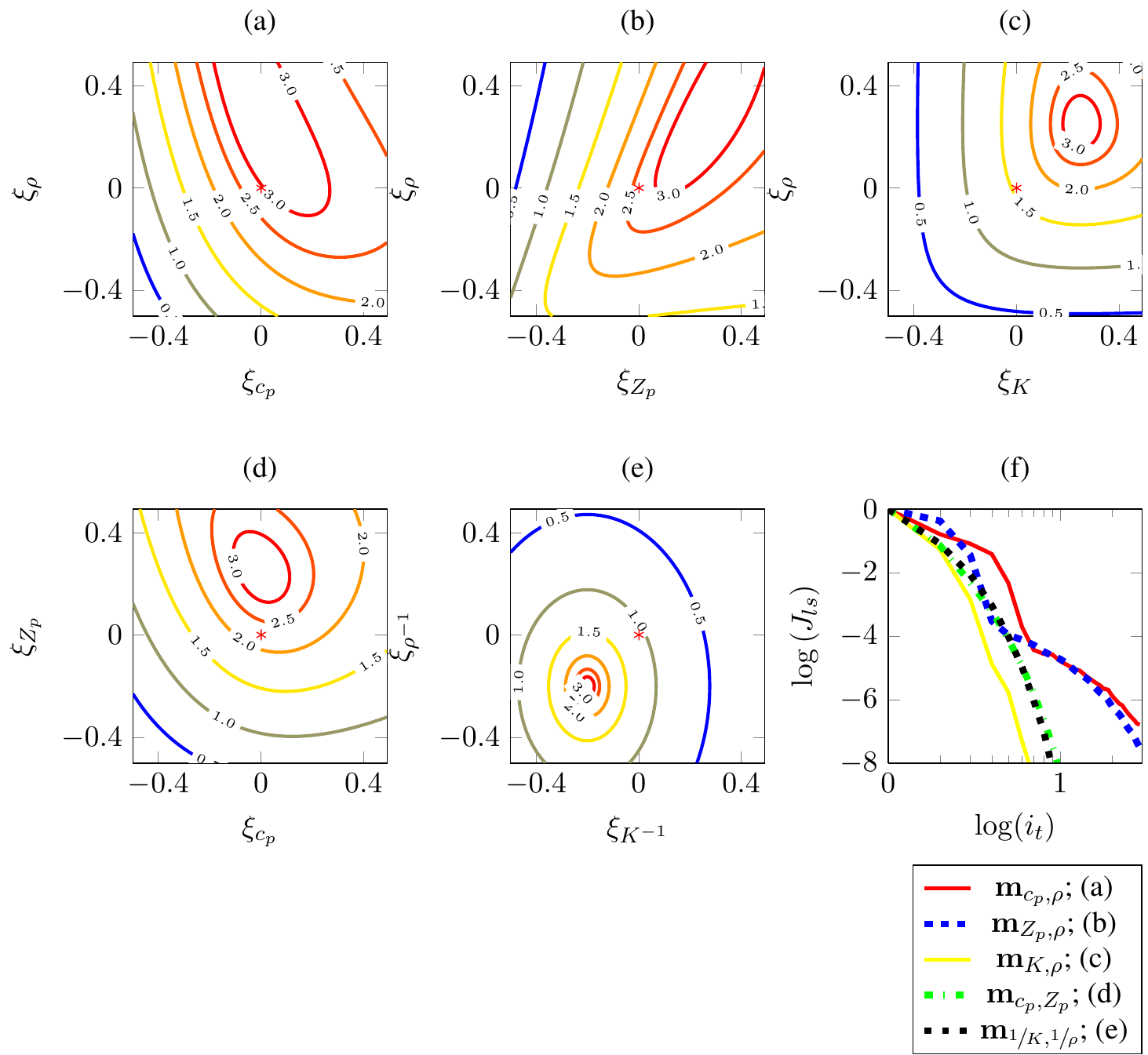}
\end{center}
\caption{
Same as Figure~\ref{fig:param_bowls_scat4_scheme1}, except for adopting scheme II, with the Born approximation and
non-linear re-parameterization.
The least-squares misfit plot,
similar to that of Figure~\ref{fig:param_bowls_scat1_scheme2}, 
shows that 
the 
suggestions of the point-scatterer analysis are 
valid.
}
\label{fig:param_bowls_scat4_scheme2}
\end{figure}

\begin{figure}
\begin{center}
\includegraphics[width=\textwidth]{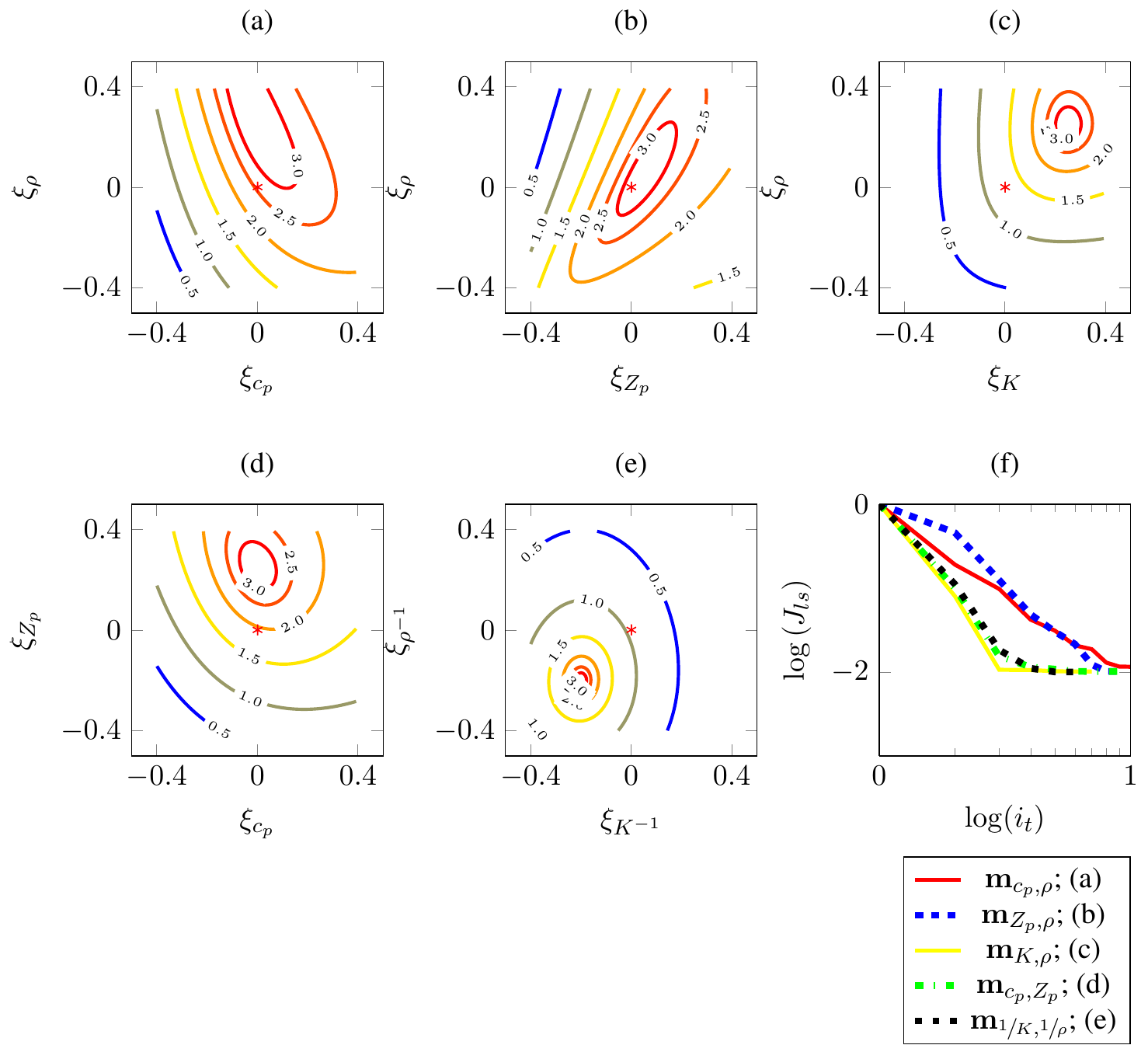}
\end{center}
\caption{
Same as Figure~\ref{fig:param_bowls_scat4_scheme1}, except for adopting scheme III, with full-waveform modelling and inversion.
It can be seen that the 
shapes of the error bowls and the relative convergence rates of different parameterization choices
are similar to that of Figure~\ref{fig:param_bowls_scat4_scheme2}.
%
%\wrmrk{why no convergence?} \prmrk{I don't understand why maybe it has to do with my finite-difference code using single-precision}
The least-squares misfit plot 
shows that 
the 
suggestions of the point-scatterer analysis are 
valid.
}
\label{fig:param_bowls_scat4_scheme3}
\end{figure}

Error-bowl analysis is performed by determining and inspecting the 
logarithmic contours of the least-squares misfit as a function of 
the two subsurface parameters.
%\subsection{Point-shaped Scatterer}
%
Reconstruction results of the point-shaped scatterer (iv) with $\mR$-only contrast 
adopting the modelling and 
inversion schemes I, II and II are given in Figures~\ref{fig:param_bowls_scat4_scheme1}, 
\ref{fig:param_bowls_scat4_scheme2} and \ref{fig:param_bowls_scat4_scheme3}, respectively.
It can be observed that 
the \velipx-, \lxxrho- and \ilxirh-parameterizations have 
faster convergence than the others, irrespective of the scheme adopted.
Point-scatterer 
analysis suggests that 
the \ilxirh and \lxxrho 
parameterization choices 
are equivalent and have 
the fastest convergence for the chosen circular acquisition geometry.
It can be seen that these 
suggestions are valid in this case.

\begin{figure}
\begin{center}
	\includegraphics[width=\textwidth]{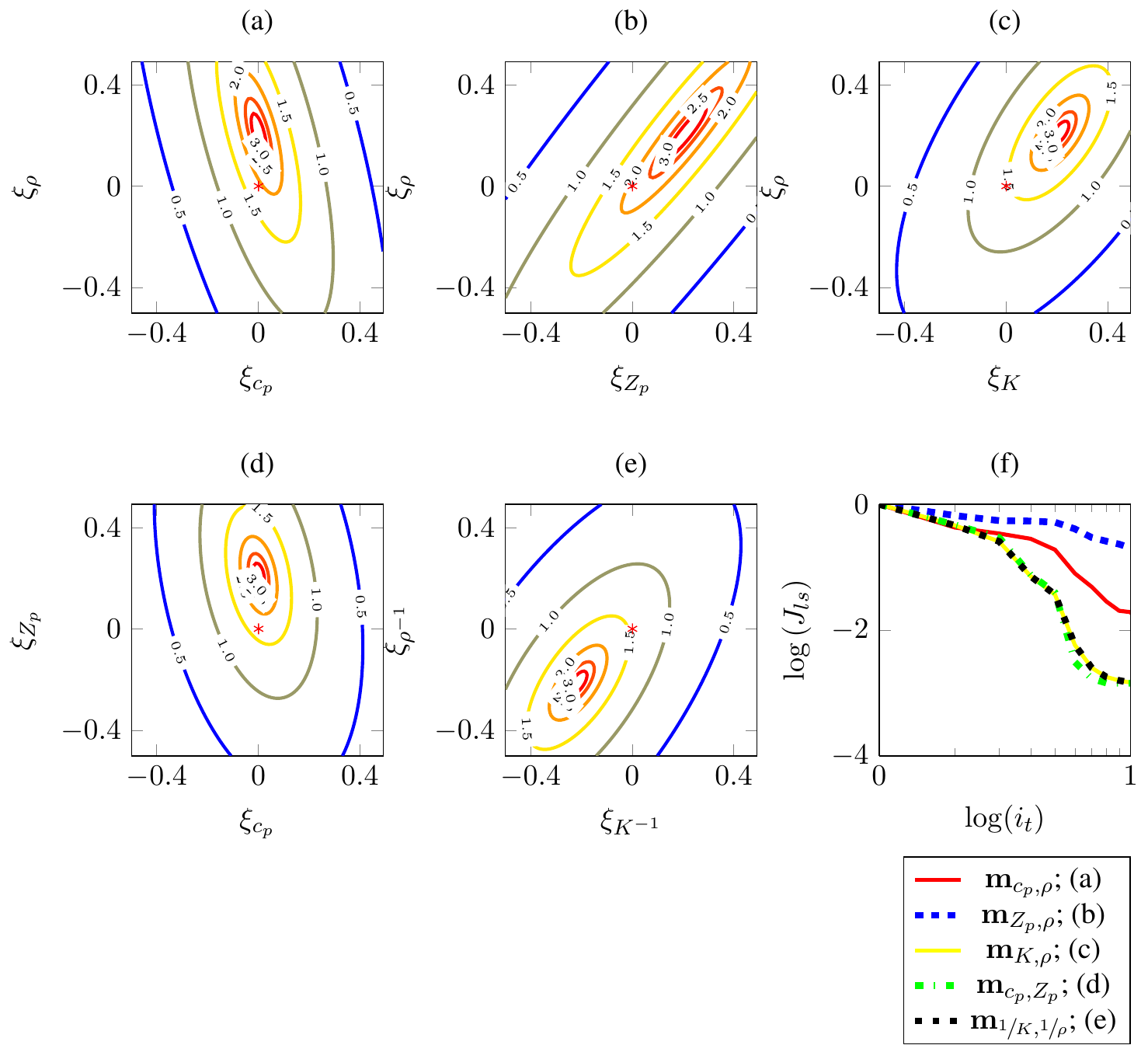}
\end{center}
\caption{
Same as Figure~\ref{fig:param_bowls_scat4_scheme1}, except for the Gaussian-shaped $\mR$-only scatterer (v).
We observe that the errors bowls for all the parameterization choices are more elliptical compared to that of Figure~\ref{fig:param_bowls_scat4_scheme1}.
The least-squares misfit plot shows that the 
suggestions of the 
point-scatterer analysis are valid. 
The rate of convergence in the case of \velipx-, \lxxrho- and \ilxirh-parameterizations 
is higher as their corresponding error bowls are more circular compared to the others.
}
\label{fig:param_bowls_scat5_scheme1}
\end{figure}

\begin{figure}
\begin{center}
	\includegraphics[width=\textwidth]{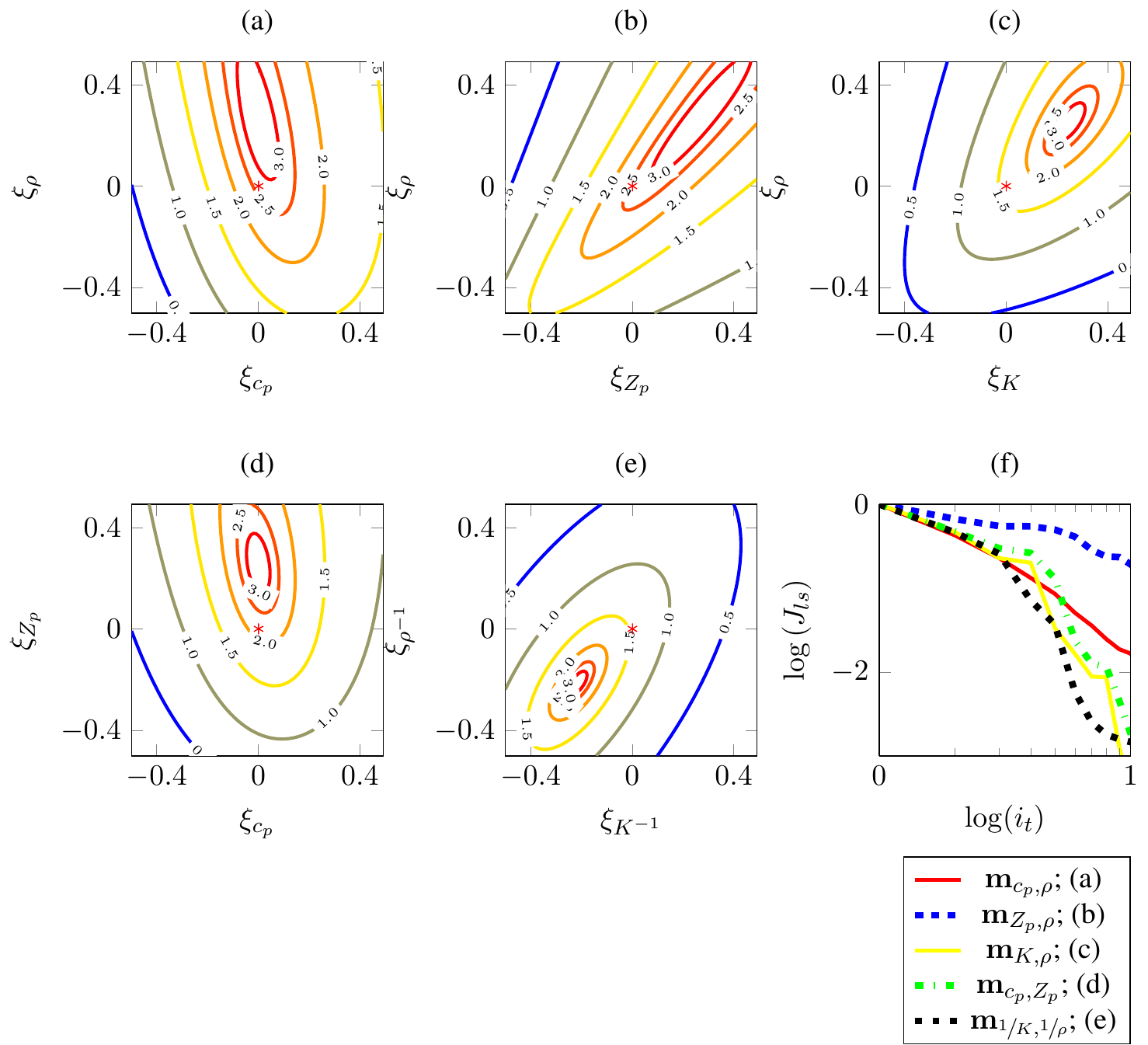}
\end{center}
\caption{
Same as Figure~\ref{fig:param_bowls_scat4_scheme1}, except for scatterer (v) and adopting scheme II, with the Born approximation and non-linear re-parameterization.
}
\label{fig:param_bowls_scat5_scheme2}
\end{figure}

\begin{figure}
\begin{center}
	\includegraphics[width=\textwidth]{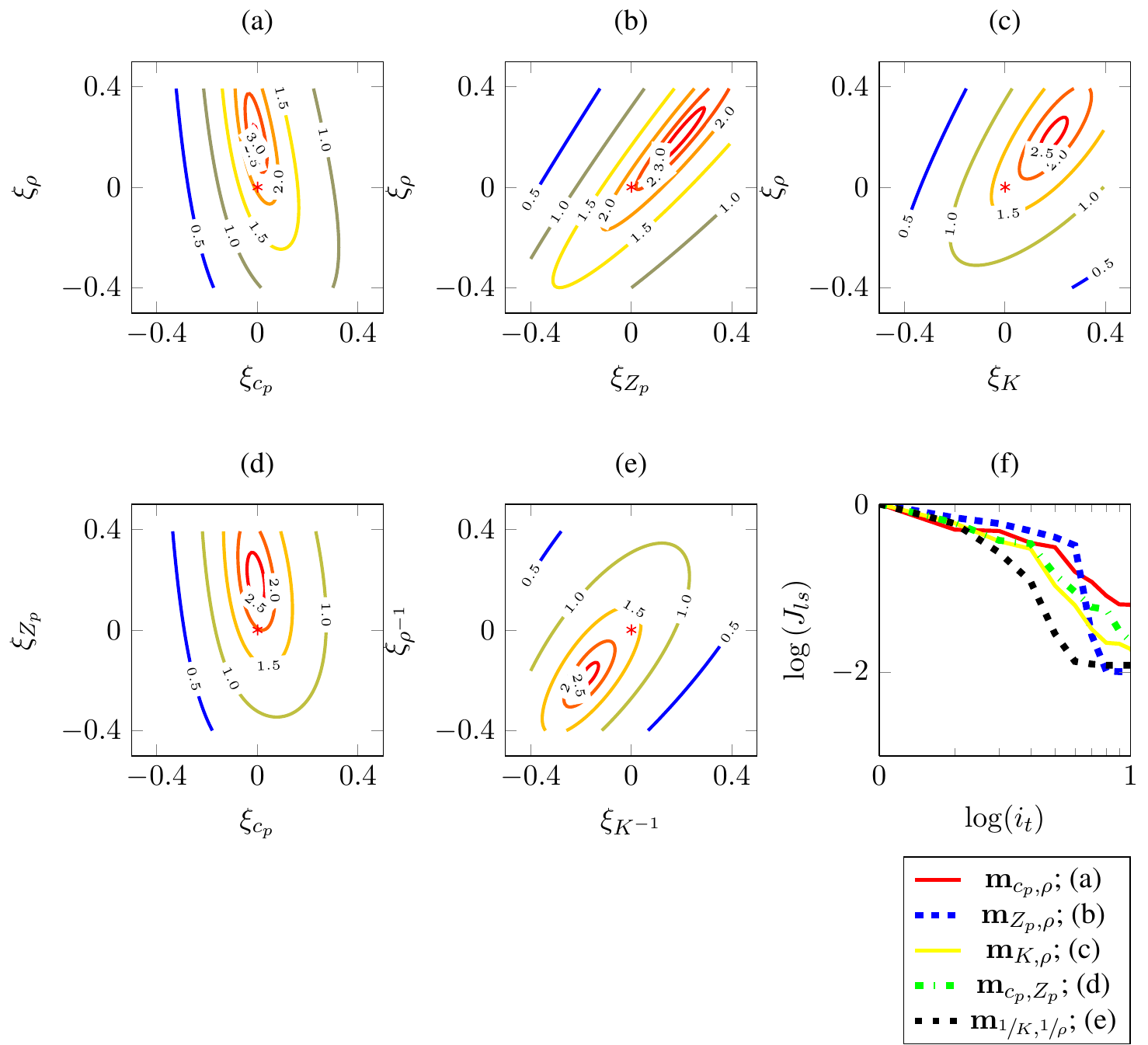}
\end{center}
\caption{
Same as Figure~\ref{fig:param_bowls_scat4_scheme1}, except for scatterer (v) and adopting scheme III, with full-waveform modelling and inversion.
Parameterization using \ilxirh has the best convergence rate. 
}
\label{fig:param_bowls_scat5_scheme3}
\end{figure}

	Reconstruction results 
of the sub-wavelength Gaussian-shaped 
scatterer (v) with $\mR$-only contrast 
adopting the modelling and 
inversion schemes I, II and II are given in Figures~\ref{fig:param_bowls_scat5_scheme1}, 
\ref{fig:param_bowls_scat5_scheme2} and \ref{fig:param_bowls_scat5_scheme3}, respectively.
In the case of schemes II and III,
the error bowls have different 
ellipticities and orientations compared to scheme I.
Here,
the \ilxirh-parameterization has the fastest convergence 
and \velrho- and \ipxrho- parameterizations 
have a slower convergence.
We observe that  
the suggestions of 
the point-scatterer analysis are 
only valid when scheme I is adopted.

%

%\bsp % ``This paper has been produced using the Blackwell
     %   Publishing GJI \LaTeXe\ class file.''

\label{lastpage}

\end{document}